\documentclass[english, 11pt]{article}
\usepackage[T1]{fontenc}
\usepackage[latin9]{inputenc}
\usepackage{geometry}
\usepackage{float}
\usepackage{graphicx}
\usepackage{babel}
\usepackage{amssymb}
\usepackage{amsmath}
\usepackage{appendix}
\usepackage{color}
\usepackage{multirow}
\usepackage{longtable}
\usepackage{booktabs}
\usepackage{array}
\usepackage{tablefootnote}
\usepackage{url}
\usepackage{graphicx} 
\usepackage{subcaption}
\usepackage{comment}
\usepackage{textcomp}
\usepackage{xcolor}   
\usepackage{amsthm}
\newcommand{\PreserveBackslash}[1]{\let\temp=\\#1\let\\=\temp}
\newcolumntype{C}[1]{>{\PreserveBackslash\centering}p{#1}}
\newcolumntype{R}[1]{>{\PreserveBackslash\raggedleft}p{#1}}
\newcolumntype{L}[1]{>{\PreserveBackslash\raggedright}p{#1}}

\newcommand{\beginsupplement}{%
        \setcounter{table}{0}
        \renewcommand{\thetable}{S\arabic{table}}%
        \setcounter{figure}{0}
        \renewcommand{\thefigure}{S\arabic{figure}}%
        \renewcommand{\thesection}{S}%
     }

\newcolumntype{P}[1]{>{\centering\arraybackslash}p{#1}}
\newcolumntype{M}[1]{>{\centering\arraybackslash}m{#1}}

\begin{document}
\title{Practical considerations for Gaussian Process modeling for causal inference quasi-experimental studies with panel data}
\author{Sofia L. Vega and Rachel C. Nethery}

\maketitle

\begin{abstract}
Estimating causal effects in quasi-experiments with spatio-temporal panel data often requires adjusting for unmeasured confounding that varies across space and time. Gaussian Processes (GPs) offer a flexible, nonparametric modeling approach that can account for such complex dependencies through carefully chosen covariance kernels. In this paper, we provide a practical and interpretable framework for applying GPs to causal inference in panel data settings. We demonstrate how GPs generalize popular methods such as synthetic control and vertical regression, and we show that the GP posterior mean can be represented as a weighted average of observed outcomes, where the weights reflect spatial and temporal similarity. To support applied use, we explore how different kernel choices impact both estimation performance and interpretability, offering guidance for selecting between separable and nonseparable kernels. Through simulations and application to Hurricane Katrina mortality data, we illustrate how GP models can be used to estimate counterfactual outcomes and quantify treatment effects. All code and materials are made publicly available to support reproducibility and encourage adoption. Our results suggest that GPs are a promising and interpretable tool for addressing unmeasured spatio-temporal confounding in quasi-experimental studies.
\end{abstract}

\section{Introduction}

Quantifying effects of past policies, interventions, and other exogenous shocks on health outcomes is critical for designing and refining health-protective programs. Such interventions or shocks can often be framed as quasi-experiments, and if longitudinal or ``panel'' data are available both before and after the shock, then its effects on health can be estimated using rigorous quasi-experimental panel data methods for causal inference. These methods leverage the panel data to estimate counterfactual health outcomes, i.e., outcomes that would have occurred in the absence of the intervention/shock, and compare those estimates to the outcomes actually observed in the post-shock period.

Gaussian Processes (GPs) have emerged as a promising approach for causal inference, offering flexibility in modeling complex, non-linear relationships within data \cite{li2021multi}. GPs are flexible models that make predictions by learning smooth patterns in the data, allowing for the incorporation of prior knowledge and uncertainty in predictions. They can also easily accommodate a variety of dependence structures in the data. At the core of any GP model is the kernel function, which defines the covariance structure of the latent process. The kernel encodes assumptions about how outcomes are expected to co-vary across units and time points, effectively controlling how much information the model borrows between observations. In the context of causal inference, in both quasi-experimental and traditional observational data settings, GPs have been trained on the observed data and then used to predict counterfactual outcomes \cite{ren2021bayesian, ben2023estimating}, providing a probabilistic framework that can accommodate various sources of uncertainty.

Recent work by Ben-Michael et al. proposed GPs for causal inference in quasi-experiments with panel data \cite{ben2023estimating}. Although promising, their uptake for quasi-experimental causal inference tasks may be hindered because GPs are ``black-boxes'' and their modeling approach and outputs are not easily interpretable. For many, it can be difficult to understand what GPs are doing under the hood, what assumptions they imply, or how modeling decisions (like the choice of kernel and tuning parameters) impact results. These concerns are particularly important in causal inference settings, where users must defend the plausibility of strong identifying assumptions, and in panel data contexts where data often have complex spatio-temporal structures.
 
To address these limitations, here we provide a practical guide to the use of GPs for conducting causal inference in quasi-experimental settings with panel data. This paper seeks to increase the accessibility and interpretability of GPs and to enable more rigorous applications of them in real panel data, where it may be necessary to empirically assess assumptions and account for complex spatio-temporal dependence. We aim to (1) clarify the requisite causal identifying assumptions, (2) increase interpretability of GPs by showing connections between GPs and weighting methods for causal inference with quasi-experiments and providing illustrative visualizations (3) conduct simulation studies to illustrate the impacts of kernel specifications in panel data, and (4) suggest diagnostics to help with kernel selection in practice, enabling researchers to make more informed modeling decisions. 

\section{Methods}

\subsection{Motivating Data}\label{sec:data}

Hurricane Katrina, one of the most devastating tropical storms in U.S. history, caused a significant public health crisis in New Orleans and the surrounding areas. Katrina made landfall on August 29, 2005, causing widespread infrastructure disruption, prolonged displacement, and elevated mortality across the Gulf Coast \cite{brunkard_hurricane_2008}. It represents an exogenous shock and so can be framed as a quasi-experiment for the purposes of estimating its impact on health outcomes. Here we seek to estimate the impacts of Hurricane Katrina on mortality among older adults in the U.S. Medicare program.

We define exposure to Hurricane Katrina at the county level. Following the exposure classification approach in Nethery et al. \cite{nethery_integrated_2023}, a county is classified as Katrina-exposed/treated if it experienced sustained gale-force winds ($\geq17.5$ m/s) from the storm at its population mean center, based on windspeed data from the \texttt{hurricaneexposuredata} R package \cite{ anderson_assessing_2020}. 96 counties are classified as treated. 81 untreated eastern U.S. counties that were located within 150 miles of a treated county are eligible for inclusion in our analyses as controls \cite{yan_tropical_2021}.

Our outcomes of interest are county-level counts of all-cause mortality among the Medicare population during the period surrounding Katrina's landfall. These data have been described in detail in previous work \cite{nethery_integrated_2023}. The study period is defined as a 140 day window surrounding Katrina's first approach to the U.S., beginning 129 days prior to its first approach and ending 11 days after. From the 177 eligible counties described above (96 treated and 81 controls), we further exclude counties with fewer than 100 Medicare enrollees or fewer than five deaths during the study period, following Nethery et al. \cite{nethery_integrated_2023}. 
This results in 150 counties, 78 treated and 72 controls (Figure \ref{fig::map}). The final panel dataset is constructed by temporally aggregating the daily mortality counts to obtain two-week mortality counts for each of the ten consecutive two-week intervals in the study period for each county. This is done to mitigate noise/instability in daily county mortality trends. 

The control counties are considered untreated throughout the entire study period. For treated counties, the first nine two-week intervals are considered ``pre-treatment'' periods and the final two-week period is the treatment period. Our aim is to build a model to learn from the trends in mortality in control counties and in treated counties pre-treatment to estimate the mortality that would have occurred in treated counties during the treatment period but in the absence of Katrina. We then compare this estimated counterfactual mortality in the absence of the storm to the observed mortality in treated counties during treatment to estimate the storm's impacts on mortality.

\subsection{Notation and Estimand}

Consider unit-time panel data including both treated and untreated units and time periods both before and after treatment. Let units be indexed by $i=1,...,N$ and time periods be indexed by $t=1,...,T$. Further, let $N_1$ denote the number of treated units. Each treated unit adopts treatment at the same time $T^*\in\{1,\ldots,T\}$.  We then define $T_0 \;=\; T^* - 1$ as the (common) number of pre-treatment periods. It may be useful to think of the outcome data as organized into an $N \times T$ panel matrix, with units as rows and time periods as columns. The observed outcome for unit $i$ at time $t$ is denoted as $Y_{it}$. Define $D_{it}\in\{0,1\}$ as a treatment indicator for each unit and time period. We further let $\textbf{X} \in \mathbb{R}^{N \times T \times p} $ denote a set of observed covariates, where $ p $ is the number of covariates and each $\textbf{X}_{it} \in \mathbb{R}^p $ corresponds to unit $ i $ at time $t$.

Under the standard stable unit treatment value assumption (SUTVA), which is formalized below, we situate our approach within the potential outcomes framework \cite{Rubin1974}. Let $Y_{it}(1)$ and $Y_{it}(0)$ denote the outcomes that would have been observed in unit $i$ at time $t$ under treatment and control, respectively. Following convention in quasi-experimental studies with panel data, we seek to estimate the 
average treatment effect on the treated (ATT), defined as 
$$
\mathrm{ATT} = \mathbb{E} \left[ Y_{it}(1) - Y_{it}(0) \,\Big|\, D_{it}=1\right].
$$
For $D_{it}=1$, $Y_{it}=Y_{it}(1)$, indicating that the observed outcome equals the potential outcome under treatment for treated units at post-treatment times. Conversely, for $D_{it}=0$, $Y_{it}=Y_{it}(0)$. As $Y_{it}(0)$ is unobserved for treated units at post-treatment times, the goal of the methods introduced in this paper is to estimate $Y_{it}(0)$ for these units, which is then used to estimate the ATT.

\subsection{Identification Assumptions for GP Framework}

We outline here the explicit assumptions required to identify causal effects in quasi-experimental panel data settings using a GP approach, which follow closely those presented by Pang et al. \cite{pang_bayesian_2022}. These assumptions integrate standard causal inference conditions and Gaussian Process modeling considerations, clearly defining the conditions under which causal interpretations from our GP approach are valid.

\begin{enumerate}

\item \textbf{Stable Unit Treatment Value Assumption (SUTVA)}: 
We assume SUTVA \cite{rubin1980randomization}, which consists of two key components:
\begin{itemize}
    \item[(a)] There is only one version of the treatment.
    \item[(b)] The potential outcome for any unit is unaffected by the treatment assignments of other units.
\end{itemize}

Formally, this implies:
$$
Y_{it} = D_{it}Y_{it}(1) + (1 - D_{it})Y_{it}(0).
$$

\item \textbf{No Anticipation of Treatment}:
Future treatment assignments do not affect current or past outcomes. Thus, potential outcomes at time $t$ depend solely on treatment statuses at or before time $t$.

\item \textbf{Treatment Path Determined by Initial Adoption}:
Each unit's entire historical treatment vector is determined completely by its initial treatment adoption time. Specifically, treated units initiate treatment at time $T^*$ and remain continuously treated until the end of the observation period ($T$). Formally:
$$
D_{it} = \mathbb{I}(t \geq T^*).
$$

\item \textbf{Individualistic Assignment and Positivity}:
We assume that treatment assignment is individualistic, meaning that the timing of treatment for any unit $i$ depends only on its own observed covariates and potential outcomes under control, and each unit has a nonzero probability of belonging to either the treated or control group.  
  
Let $D_i=\mathbb{I}(\exists t : D_{it}=1)\in\{0,1\}$ be the indicator that unit $i$ is ever treated (so that $D_{it}=\mathbb{I}(t\ge T^*)\,D_i$),  $\mathbf{X}_{i\scriptscriptstyle\bullet}=(X_{i1},\dots,X_{iT})$ be the history of covariates for unit $i$, and $Y_{i\scriptscriptstyle\bullet}(0)=(Y_{i1}(0),\dots,Y_{iT}(0))$ be the history of its control potential outcomes.  

Formally, we require:
$$
\Pr(\mathbf{D}\mid X,\,Y(0))
\;=\;\prod_{i=1}^N
  \Pr\bigl(D_{i}\mid \textbf{X}_{i\scriptscriptstyle\bullet},\,Y_{i\scriptscriptstyle\bullet}(0)\bigr),
$$
and
$$
0 < \Pr\bigl(D_i=1 \mid \textbf{X}_{i\scriptscriptstyle\bullet},\,Y_{i\scriptscriptstyle\bullet}(0)\bigr) < 1
\quad\forall\,i.
$$

\item \textbf{Latent Ignorability}:
Conditional on observed covariates, $\mathbf{X}$, and a latent Gaussian process $f_{it}$ capturing unmeasured confounding, treatment assignment is independent of potential outcomes under control:
$$
Y_{it}(0) \perp D_{it} \mid f_{it}, \textbf{X}.
$$

\item \textbf{Known Covariance Structure of Latent Gaussian Process}:
Unobserved confounding factors can be represented by a Gaussian process with covariance governed by a kernel function $K$ with known structure, but possibly depending on unknown parameters:
$$
f \sim \mathcal{GP}(0, K((i,t), (i',t'))).
$$

\item \textbf{Conditional Exchangeability under Gaussian Process Priors}:
Conditional on the latent Gaussian process $f$ and observed covariates $\mathbf{X}$, the observed outcomes under control, $Y_{it}(0)$, are exchangeable enabling inference through posterior predictive distributions.

\end{enumerate}

Identification follows as shown in Pang et al. \cite{pang_bayesian_2022}. These assumptions form the foundation for valid causal inference within the GP framework. 

\subsection{Gaussian Processes for Panel Data}

We initially present the GP approach for Normally distributed outcomes, and we later show how other outcome distributions can be accommodated within this framework. Ben-Michael et al. focus their model formulation and analysis on settings with a single treated unit \cite{ben2023estimating}; yet, the GP framework can be naturally extended to handle multiple treated units. Here, everything is presented in sufficient generality to allow for an arbitrary number of both treated and control units. 

Following Ben-Michael et al. \cite{ben2023estimating}, assume the potential outcomes under control arise from the following Gaussian Process model:

\begin{equation}\label{eq:gpmain}
\begin{aligned}
    Y_{it}(0) &= \mu_0 + f_{it} + \mathbf{X}_{it}^\top \boldsymbol{\beta} + \varepsilon_{it}, \\
\mathbf{f} &\sim \mathcal{GP}\left(0, \mathbf{K}\right), \\
\varepsilon_{it} &\overset{\text{i.i.d.}}{\sim} \mathcal{N}(0, \sigma^2)
\end{aligned}
\end{equation}
where $\mu_0$ is a global intercept, $\mathbf{f}=(f_{11},...,f_{NT})$ is is the vector of model components for all units and time period with covariance matrix $\mathbf{K} \in \mathbb{R}^{NT \times NT}$, $\mathbf{X}_{it}$ is a vector of time-varying measured confounders, and $\boldsymbol{\beta}$ is a vector of their coefficient parameters. The elements of $\mathbf{K}$ are given by a kernel function, $k((i,t),(i',t'))$, characterizing the covariance between any two unit-time pairs $(i,t)$ and $(i',t')$. The kernel function typically depends on several unknown parameters, as discussed below. 

In many applications, one may wish to account for unit-specific baseline outcome levels or time-point-specific shocks (aside from the shock of interest) that are not well captured by a GP. To do this, one can simply augment $\mathbf{X}$ with dummy-indicator columns for each unit and/or each time period, effectively adding unit and/or time fixed effects. These fixed-effect terms can improve robustness when there are abrupt changes in time or persistent differences across units under control conditions that may violate the smoothness assumptions of the kernel.

Throughout this paper, we will take a fully Bayesian approach to estimation of this model. Thus, the covariance parameters are learned from the data and then values of the missing $Y_{it}(0)$ can be predicted by leveraging information from correlated units and/or time periods. Under the GP model proposed above, the vector of all potential outcomes under control, $\mathbf{Y(0)}$, has the following multivariate Normal distribution:
$$\mathbf{Y(0)}=
\left(\begin{array}{l}
\mathbf{Y(0)}^{\mathrm{mis}} \\
\mathbf{Y(0)}^{\mathrm{obs }}
\end{array}\right) \sim \operatorname{MVN}\left[\left(\begin{array}{l}
0 \\
0
\end{array}\right),\left(\begin{array}{cc}
\mathbf{K}_{\mathrm{mis }} & \mathbf{K}_{\mathrm{mis }, \mathrm{ obs }} \\
\mathbf{K}_{\mathrm{obs }, \mathrm{ mis }} & \mathbf{K}_{\mathrm{obs }}
\end{array}\right)+\sigma^2 \mathbb{I}\right]
$$ 
where $\mathbf{Y(0)}$ is partitioned into a vector of the observed outcomes under control for treated units pre-treatment ($\mathbf{Y(0)}^{\mathrm{obs }} \in \mathbb{R}^{(N-N_1)T + N_1(T_0)}$) and a vector of the missing outcomes under control for the treated units and post-treatment times ($\mathbf{Y(0)}^{\mathrm{mis }} \in \mathbb{R}^{N_1(T-T_0)}$). Similarly, the covariance matrix ($\mathbf{K}$) can be divided into matrices for the observed outcomes under control ($\mathbf{K}_{\mathrm{obs }} \in \mathbb{R}^{((N-N_1)T+(N_1)T_0) \times (N-N_1)T+(N_1)T_0)}$), the missing outcomes under control for the treated units post-treatment ($\mathbf{K}_{\mathrm{mis }} \in \mathbb{R}^{N_1(T-T_0) \times N_1(T-T_0)}$), and the covariance between the observed and missing outcomes under control ($\mathbf{K}_{\mathrm{mis }, \mathrm{ obs }} \in \mathbb{R}^{(N-N_1)T \times N_1(T-T_0)}$).

Applying standard properties of multivariate Normal distributions, the posterior predictive distribution for the missing potential outcomes under control is then given by:
$$
\begin{aligned}
\mathbf{Y(0)}^{\mathrm{mis}} \mid \mathbf{Y(0)}^{\mathrm{obs }} & \sim \operatorname{MVN}(\boldsymbol{\mu}, \boldsymbol{\Sigma}) \\
\boldsymbol{\mu} & =\mathbf{K}_{\mathrm{mis,obs }}\left(\mathbf{K}_{\mathrm{obs }}+\sigma^2 \mathbb{I}\right)^{-1} \mathbf{Y(0)}^{\mathrm{obs }} \\
\boldsymbol{\Sigma} & =\mathbf{K}_{\mathrm{mis }}-\mathbf{K}_{\mathrm{mis }, \mathrm{obs}}\left(\mathbf{K}_{\mathrm{obs}}+\sigma^2 \mathbb{I}\right)^{-1} \mathbf{K}_{\mathrm{obs}, \mathrm{mis}} .
\end{aligned}
$$ The posterior mean is thus expressed as a linear combination of the observed outcomes under control, with the coefficients of the linear combination determined by the covariances.

To obtain estimates of the treatment effect, the missing entries in $\mathbf{Y(0)}$ are sampled from the posterior predictive distribution. MCMC samples of the ATT, ATT$^{(m)}$ for $m=1,...,M$, are constructed as:
$$\mathrm{ATT}^{(m)} = \frac{1}{|\{(i,t): D_{it} = 1\}|} \sum_{(i,t): D_{it} = 1} \left[ Y_{it} - Y_{it}^{(m)}(0) \right]$$
where $Y_{it}^{(m)}(0)$ is the $m^{\mathrm{th}}$ MCMC sample of $Y_{it}(0)$. Using summaries of the MCMC samples, we can estimate $\mathrm{ATT}$ and its associated uncertainties.

\subsection{Extending to Other Outcome Distributions}

While we have thus far described the GP framework assuming Normally distributed outcomes, the model can be naturally extended to accommodate non-Gaussian data by incorporating appropriate link functions \cite{ben2023estimating}.

For count outcomes, such as mortality counts for areal units, a Poisson likelihood with a log link function can be used:
$$
Y_{it}(0) \sim \text{Poisson}(\mu_{it}), \quad \log(\mu_{it}) = f_{it} + \log(\theta_{it}),
$$
where $ \textbf{f} \sim \mathcal{GP}(0, \textbf{K}) $ is the Gaussian process and $ \theta_{it} $ is a known offset, such as a population exposure term.

For binary outcomes, such as event indicators, a Bernoulli likelihood with a logit link can be used:
$$
Y_{it}(0) \sim \text{Bernoulli}(\pi_{it}), \quad \text{logit}(\pi_{it}) = f_{it},
$$
where again $ \textbf{f} \sim \mathcal{GP}(0, \textbf{K}) $.

\section{Kernel Selection and Confounding Adjustment}

In this section, we first explain the role of the latent Gaussian process in accounting for potential unmeasured confounding, then we discuss how the choice of kernel function selected when fitting the GP model might allow for different types of unmeasured confounders to be captured. $f_{it}$ is a latent process capturing trends over space and time in the outcomes, which represents the influence of unmeasured variables that are potential confounders of the effect of treatment. That is, we assume the presence of an unmeasured, time-varying factor $ f_{it} $ that influences the potential outcomes and may also vary systematically by treatment status. Temporarily omitting observed covariates for ease of explanation, recall that under Equation~\ref{eq:gpmain}, $ Y_{it}(0) = f_{it} + \varepsilon_{it}$. To more clearly illustrate the role of $f_{it}$, further suppose the potential outcomes under treatment arise from the following model:

$$Y_{it}(1) = f_{it} + \tau_{it} + \varepsilon_{it},
$$
where  $\tau_{it}$  denotes the treatment effect and $f_{it}$ and $\varepsilon_{it}$ are as in Equation~\ref{eq:gpmain}. (Note that our approach does not fit a model for $Y_{it}(1)$, we just use this modeling assumption to elucidate the role of $f_{it}$.) In the GP model fitting, $f_{it}$ is learned from the pre-treatment and control observations via the kernel $K$, such that for $(i,t):\ D_{it}=1$ (treated units at post-treatment times),
$\widehat{Y}_{it}(0)=\widehat{f}_{it}$. Under the causal consistency assumption, for $(i,t):\ D_{it}=1$, $Y_{it}=Y_{it}(1)=f_{it} + \tau_{it} + \varepsilon_{it}$ and plugging into the ATT estimator given in the previous section
\begin{align*}
    \widehat{\mathrm{ATT}} & = \frac{1}{|\{(i,t): D_{it} = 1\}|} \sum_{(i,t): D_{it} = 1} \left[ Y_{it} - \widehat{Y}_{it}(0) \right] \\
    & = \frac{1}{|\{(i,t): D_{it} = 1\}|} \sum_{(i,t): D_{it} = 1} \left[ f_{it} + \tau_{it} + \varepsilon_{it} - \widehat{f}_{it} \right]\\
    & = \frac{1}{|\{(i,t): D_{it} = 1\}|} \left[ \sum_{(i,t): D_{it} = 1}  \tau_{it} \right] + \frac{1}{|\{(i,t): D_{it} = 1\}|} \left[\sum_{(i,t): D_{it} = 1} \varepsilon_{it} + \sum_{(i,t): D_{it} = 1} (f_{it} - \widehat{f}_{it}) \right]
\end{align*}
where the second term in the last line will approximate zero if $\widehat{f}_{it}$ is an unbiased estimator of $f_{it}$. Thus, if we obtain an unbiased estimate of the latent process $f_{it}$ from the GP model, the ATT estimate will be unbiased for the true ATT, $\frac{1}{|\{(i,t): D_{it} = 1\}|} \left[ \sum_{(i,t): D_{it} = 1}  \tau_{it} \right]$.
This clarifies how proper specification and estimation of $f_{it}$ will allow for adjustment for unmeasured confounding and unbiased estimation of the causal quantity of interest. The choice of kernel impacts the types of dependence patterns, and therefore the types of unmeasured variables, that are captured in $f_{it}$, which therefore impacts the validity and accuracy of the ATT estimates from the GP model. Prior recommendations for the kernel parameters are given in Section \ref{sec::priors}.

\subsection{Types of Covariance Structures Addressed by Kernels}

Gaussian Process kernels specify assumptions about data dependence structures, allowing models to flexibly adjust for various forms of unmeasured confounding encountered in panel data settings. For example, unmeasured temporal confounding occurs when magnitudes and trends in the outcome change systematically pre- to post-treatment, due to some unmeasured time-varying variable. Kernels explicitly structured to capture temporal auto-correlation may be most useful for mitigating bias from unmeasured confounders with smooth temporal trends. Conversely, unmeasured spatial confounding arises when outcomes in treated and control units differ systematically due to unmeasured spatially-varying variables. Kernels explicitly structured to capture spatially smooth dependencies based on geographic proximity may be most useful for mitigating bias from unmeasured confounders that vary smoothly across space \cite{reich2021review}. 

Often, complex interactions between spatial and temporal dimensions occur in real-world data. So-called ``separable'' kernels-- the primary type of kernels used in practice-- assume spatial correlations remain constant over time and temporal correlations remain constant over space, limiting their ability to capture dynamic space-time interactions. Nonseparable kernels allow the covariance between observations to depend explicitly on combined spatial and temporal distances, potentially enhancing the ability to capture complex spatio-temporal confounding patterns \cite{gneiting2002nonseparable}; however they are rarely used in practice. By aligning kernel selection with the nature of the assumed confounding structures, researchers can effectively control latent sources of bias. In this section we discuss some standard kernel choices and their implications for confounding adjustment. While many other kernel choices are possible (so long as they produce symmetric and positive-definite covariance matrices), for tractability we restrict our discussion here to three kernels, each a representative of a different class of kernels and each with different confounding adjustment implications.

\subsection{Separable Kernels}

Separable kernels are a common choice in Gaussian Process modeling due to their simplicity and computational efficiency. Separable kernels are defined by their ability to be decomposed into separate time and unit (space) covariance kernels, i.e., $ \mathbf{K} = \mathbf{K}_{\mathrm{unit}} \otimes \mathbf{K}_{\mathrm{time}}.$ This places the restriction that the covariance across units is constant in time and the covariance across time is constant across units.

\subsubsection{Separable ICM-RBF Kernel}
In their study, Ben-Michael et al. utilized a separable kernel, specifying an Intrinsic Coregionalization Model (ICM) kernel to capture across-unit dependencies and a radial basis function (RBF) kernel to capture temporal dependencies \cite{ben2023estimating}. In the ICM kernel, $ k_{\text{unit}}(i, i') =\sum_{j=1}^J \phi_{ij} \phi_{i'j}$, where the $\phi_{ij}$ are unknown parameters that can be organized into a rank-J matrix $\phi \in \mathbb{R}^{N \times J}$, $J<N$. The ICM kernel ensures identifiability by imposing this low-rank structure on the matrix of covariance parameters. Under the ICM kernel, the covariance matrix $\mathbf{K}_{\mathrm{unit}}$, can be represented as the product of $\phi$ and its transpose: $\mathbf{K}_{\mathrm{unit}} = \phi \phi^{\mathrm{T}}$. The ICM kernel allows for unstructured dependence patterns across units, as long as those patterns can be represented by a low-rank $\phi$. This low rank constraint can make interpretation of results challenging, since it is difficult to characterize the simplifications of the dependence structures in the data it is imposing. However, the low-rank assumption reduces the number of parameters to be estimated, leading to computational efficiency, especially for large datasets. The rank, $J$, is viewed as a hyperparameter, controlling the complexity and flexibility of the model.

In Ben-Michael et al's GP implementation, the temporal dependencies are captured through an RBF kernel: 

\begin{align}\label{eq:rbftime}
k_{\text{time}}(t,t') = \exp\left(-\frac{|t-t'|^2}{2l_t^2}\right).
\end{align}
In this kernel, the correlation between time points $t$ and $t'$ decreases as their distance increases. The parameter $l_t$ controls the temporal length scale, with larger values allowing longer-range dependencies. In the Bayesian paradigm, a prior can be placed on $l_t$ and it can be learned from the data.

Although this approach reduces complexity, the linear kernel has limited expressiveness and lacks interpretability, making it difficult to understand exactly what types of unmeasured confounding it might be picking up.

\subsubsection{Separable RBF-RBF Kernel}
Instead of using the ICM kernel to capture across-unit dependence, one could consider using an explicitly spatial kernel for the unit kernel, i.e., a kernel that is structured to allow for greater dependence between units that are closer together in space. This restriction that the dependence should have a spatial structure can be thought of as an alternative constraint to the low-rank constraint of the ICM kernel, which will lead to greater interpretability in many instances. The spatial RBF kernel is a common choice for capturing spatial dependencies. Let $s_i$ represent the the spatial coordinates for unit $i$, then:
$$
k_{\text{unit}}(i,i') = \exp\left(-\frac{\|s_i-s_{i'}\|^2}{2l_s^2}\right), 
$$

where $ l_s $ is a length-scale parameter that controls the smoothness of the function. The RBF kernel measures the similarity between spatial locations based on their Euclidean distance, leading to a smooth and continuous representation of spatial dependencies. Thus, it may improve accuracy relative to the ICM kernel in settings where the unmeasured confounders are believed to vary smoothly over space. Additionally, the RBF kernel can be more computationally demanding than the ICM, especially for large datasets. To obtain a separable space-time kernel that captures both spatially smooth and temporally smooth patterns, the RBF unit kernel can be combined with an RBF time kernel as given in Equation~\ref{eq:rbftime}.

\subsection{Nonseparable Space-Time Kernels}

While separable kernels like the ICM kernel and the RBF kernel are useful, they can be limiting when modeling complex spatio-temporal dependencies. Nonseparable kernels offer a more flexible approach by allowing the covariance structure to vary across both space and time. 

In practice, it's not always true that the covariance remains constant across units over time or across time for all units. To enable a more dynamic modeling of space-time interaction, one can instead utilize a class of nonseparable space-time covariance functions described in Gneiting et al. \cite{gneiting2002nonseparable}. In contrast to separable kernels, non-separable kernels cannot be decomposed into the the kronecker product of unit and time kernels. This complicates characterization of the spatial and temporal dependence structures allowed. 

Letting $\varphi(z),z\geq 0$ be any completely monotone function and $\psi(z), z \geq 0$ be any positive function with a completely monotone derivative, then the Gneiting nonseparable covariance has the form:
$$
k((i,t), (i',t'))) = \frac{\tau^2}{\psi(|t-t'|^2)}\varphi\left(\frac{||s_i-s_{i'}||^2}{\psi(|t-t'|^2)} \right)
$$

Following Gneiting et al., if we let $\varphi(z)=\exp \left(-l_s z^\gamma\right)$ with $ l_s>0$ and $0<\gamma \leq 1$ and $\psi(z)=\left(l_t z^\alpha+1\right)^\eta$ with $ l_t>0,0<\alpha \leq 1$, and $0 \leq \eta \leq 1$, then
\begin{equation}
k((i,t), (i',t')))=\frac{\tau^2}{\left(\frac{1}{l_t}|t-t'|^{2 \alpha}+1\right)^{\eta}} \exp \left(-\frac{\frac{1}{l_s}\|s_i-s_{i'}\|^{2 \gamma}}{\left(\frac{1}{l_t}|t-t'|^{2 \alpha}+1\right)^{\eta \gamma}}\right)
\end{equation}

where $\tau^2$ is the variance of the spatio-temporal process \cite{gneiting2002nonseparable}. Here, $\eta$ represents the strength of the space-time interaction. When $\eta = 0$, there is no space-time interaction and the resulting kernel is separable. As $\eta$ increases, the interaction between space and time increases, causing spatial correlations at temporal lags other than zero to decline at a progressively slower rate \cite{gneiting2002nonseparable}. In the Gneiting kernel, the two scaling (or length-scale) parameters, $l_s$ and $l_t$, control how quickly correlations decay with spatial distance and temporal lag, respectively, while the smoothing exponents of space and time, $\gamma$ and $\alpha$, shape the ``sharpness'' or roughness of that decay at small lags. To see how these parameters change the value of the covariance, see Figure \ref{fig:gneiting_param}. This flexible structure in spatial and temporal patterns enabled by nonseparable kernels may be more plausible for capturing many types of unmeasured confounders in real data applications. However, they are substantially more computationally demanding than separable kernels, which may limit their usability in many applications.

\subsection{Connecting GPs to Existing Methods: Weighting and Kriging Representation}\label{sec:weights}

Ben-Michael et al. show that, for arbitrary kernel specifications, the posterior predictive mean of a GP can be expressed as a linear combination of the observed outcomes:

$$
\widehat{Y}_{it}(0) = \sum_{(i',t'):D_{i't'}=0} \hat{w}_{i't'}^{(i,t)} Y_{i't'},
$$
where $\hat{w}_{i't'}^{(i,t)}$ are posterior mean unit-time weights, derived from the kernel function $k((i,t), (i',t'))$ and variance $\sigma^2$ according to:
$$
\hat{w}_{i't'}^{(i,t)} = k((i,t), (i',t')) \left[k((i',t'),(i',t')) + \sigma^2 I\right]^{-1}
$$ \cite{ben2023estimating}.

When assuming the structure of the kernel is separable, Ben-Michael et al. show the posterior predictive weights can decompose into separate weights for units and time periods:
$$
\widehat{Y}_{it}(0) = \sum_{(i',t'):D_{i't'}=0} \hat{w}_{ii'}^{(\text{unit})} \hat{w}_{tt'}^{(\text{time})} Y_{i't'}
$$ where $\hat{w}_{i}^{(\text{unit})} \in \mathbb{R}^D$ is a vector of unit-level weights and $\hat{w}_{t}^{(\text{time})} \in \mathbb{R}^T$ is a vector of time-period weights, given by the following closed form \cite{ben2023estimating}:
$$
\hat{w}_{i}^{(\text{unit})} \otimes \hat{w}_{t}^{(\text{time})} = (\textbf{K}_{\mathrm{unit}}\otimes  \textbf{K}_{\mathrm{time}} + \sigma^2I)^{-1}(k_{\mathrm{unit}}(i,\cdot) \otimes k_{\mathrm{time}}(t,\cdot)).
$$

If we assume $\mathbf{K}_{\mathrm{time}} = I$ for any arbitrary $\mathbf{K}_{\mathrm{unit}}$, the GP model removes temporal smoothing and assigns weights solely based on unit-level similarity. This special case results in time-invariant unit weights applied independently at each post-treatment time point, structurally mirroring the so-called vertical regression approach commonly used in quasi-experimental studies. In vertical regression, counterfactual outcomes are estimated at each time point using a fixed weighted average of control units, without modeling temporal correlation. This structure also closely resembles Synthetic Control Methods (SCM), which estimate unit-level weights to match pre-treatment trends of treated units. However, unlike SCM, which derives weights through constrained optimization, GP derives weights from the posterior distribution of a Gaussian Process using a unit-level kernel. While SCM enforces constraints like non-negativity and convexity, GP weights can be unconstrained and reflect smoothness assumptions encoded in the kernel. Thus, vertical regression can be seen as a special case of GP under an identity time kernel, and GP provides a broader framework that connects and extends both vertical regression and SCM through kernel-based smoothing and uncertainty quantification.

Importantly, when explicitly encoding spatially and/or temporally smooth dependence structures through kernels such as the RBF, the resulting weights directly reflect spatial or temporal distances. Spatial weights $ \hat{w}_{ii'}^{(\text{unit})} $ prioritize outcomes from nearby units, while temporal weights $ \hat{w}_{tt'}^{(\text{time})} $ discount the influence of more distant time periods. This structure directly extends the SCM approach by incorporating explicit spatial correlation and temporal smoothing, thus generalizing SCM to scenarios with continuous and structured spatio-temporal effects. Alternatively, for nonseparable kernels, we can't simplify or decompose the form of the weights any further from what's shown in Section 3.3, however the weights will have an explicit spatial and temporal structure.

This weighting representation closely parallels kriging, a popular geostatistical interpolation technique. Kriging is a method for spatial interpolation that provides the best linear unbiased prediction of the intermediate values. Although not commonly applied in quasi-experimental studies, kriging can be effectively adapted for use in quasi-experiments to study treatment effects by interpolating the observed outcomes under control across space and time to estimate missing counterfactual outcomes under control.

Specifically, the counterfactual predictions from a kriging approach would be \cite{suryasentana2023demystifying}:
$$
\mathbf{Y(0)}^{\mathrm{mis}} = \tilde{\mathbf{K}}_{\mathrm{mis,obs}}(\tilde{\mathbf{K}}_\mathrm{obs}+\sigma^2\mathbb{I})^{-1} \mathbf{Y}_{\mathrm{obs}}
$$ 
which is equivalent to the mean of the posterior predictive distribution of the GP model shown in Section 2.4 when covariance specifications are equivalent. Here, $ \tilde{\mathbf{K}}_{\text{obs}}$ and $ \tilde{\mathbf{K}}_{\text{mis,obs}} $ represent the covariance matrices under the kriging model, which coincide with the corresponding blocks of $K$ in the GP framework when the same kernel function is used. Similar to GPs, weights are influenced by the kernel and can be either separable or nonseparable. This highlights that GPs effectively perform kriging over untreated observations to recover counterfactual outcomes in quasi-experimental settings.

By framing counterfactual estimation as a weighted average where weights are derived from flexible kernel structure, GPs bridge the methodological gap between traditional causal inference tools such as SCM, which focus on weighting similar control units, and kriging, which emphasizes spatial interpolation. The GP framework enhances these existing methods by explicitly learning weights through flexible kernels and capturing the complex dynamics of space-time confounding. This unified view clarifies the methodological connections and broadens the applicability of GPs as a robust tool for causal inference in quasi-experiments. It also enhances interpretability by linking the GP framework to simpler and more familiar methods like SCM, where the role of weights and their relationship to observed data are more transparent.

\section{Kernel Selection in Practice}

The previous section suggests that, when possible, the choice of kernel should be informed by substantive knowledge regarding the structures of anticipated unmeasured confounding variables. Thus, it is important that users have a \textit{practical} understanding of the types of spatio-temporal patterns that could be handled by each of the kernel specifications above. In this section, we simulate data and provide illustrations and diagnostics to facilitate such practical considerations and assist users in selecting kernels appropriate for their real panel data applications.

\subsection{Illustrating Kernel Properties}
We simulated data across a simplified spatial-temporal setup involving a 7$\times$7 spatial grid and 15 time points. 
Observations were generated from the model in Equation~\ref{eq:gpmain}, without observed covariates, under each of the three different kernel specifications described in the previous section. We consider 4 random units at time points 9-15 as treated. For simplicity of illustration and to focus squarely on kernel-driven behavior, none of the simulations below (or in Section \ref{sec::gpsim}) include any observed covariates or fixed effects in $\mathbf{X}$.

\begin{figure}[H]
    \centering
    \includegraphics[width=1\linewidth]{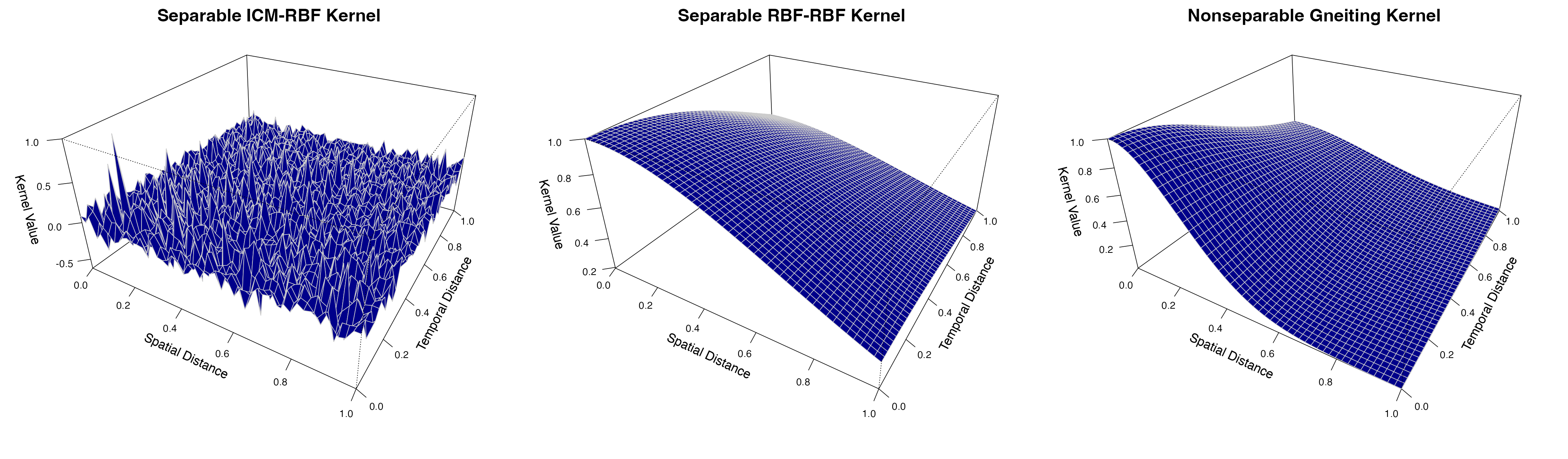}
    \caption{Kernel structure for three spatio-temporal covariance functions.
Each panel shows the kernel surface as a function of space and time for a separable ICM-RBF kernel (left), a separable RBF-RBF spatio-temporal kernel (middle), and a nonseparable Gneiting spatio-temporal kernel (right). For both separable kernels, we use a spatial lengthscale of 0.7, a temporal lengthscale of 0.9, and marginal variance $\tau^2 = 1$. For the nonseparable Gneiting kernel, we set the spatial lengthscale to 1/8 and temporal lengthscale to 1/1.75 to exaggerate space-time interaction effects for visualization purposes.}
    \label{fig:3d_kernel}
\end{figure}

Figure \ref{fig:3d_kernel} shows the kernel surfaces, i.e., the value of $k()$ as a function of the spatial and temporal distance between units. This illustrates how the covariance functions differ in terms of decay patterns across space and time under different kernel specifications. The nonseparable kernel reflects linked spatio-temporal decay as shown by the twisted contours where spatial and temporal distances are interacting multiplicatively. The separable RBF-RBF kernel's assumption of independent spatial and temporal decay is visible in the plot because, at any fixed spatial location (such as spatial distance = 0), the kernel value decreases smoothly along the time axis, while at any fixed time, the kernel value decreases smoothly with spatial distance, with no interaction or twisting between the two axes. The ICM-RBF kernel does not incorporate spatial distance directly but instead models correlations between units through shared latent processes, leading to spatially unstructured kernel surfaces. Here, the temporal RBF decay may appear less prominent due to the dominant influence of the ICM's unit-level correlations in the visualization.

As discussed in Section \ref{sec:weights}, the estimated counterfactual can be represented as a linear combination of the observed outcomes under control. Following the terminology used in the SCM literature, we call the weights in this linear combination `donor weights', which can be used to assess the degree to which each control unit/time point (donor) contributes information to the counterfactual outcome of the treated unit. The GP model defines a set of true underlying weights, determined by the specified covariance structure. In practice, these weights can be estimated from observed data by fitting the model, yielding approximation of the true weights. In this section, we visualize the true donor weights on simulated data to gain insight on how these kernels are working in practice.

\begin{figure}[H]
    \centering
    \includegraphics[width=1\linewidth]{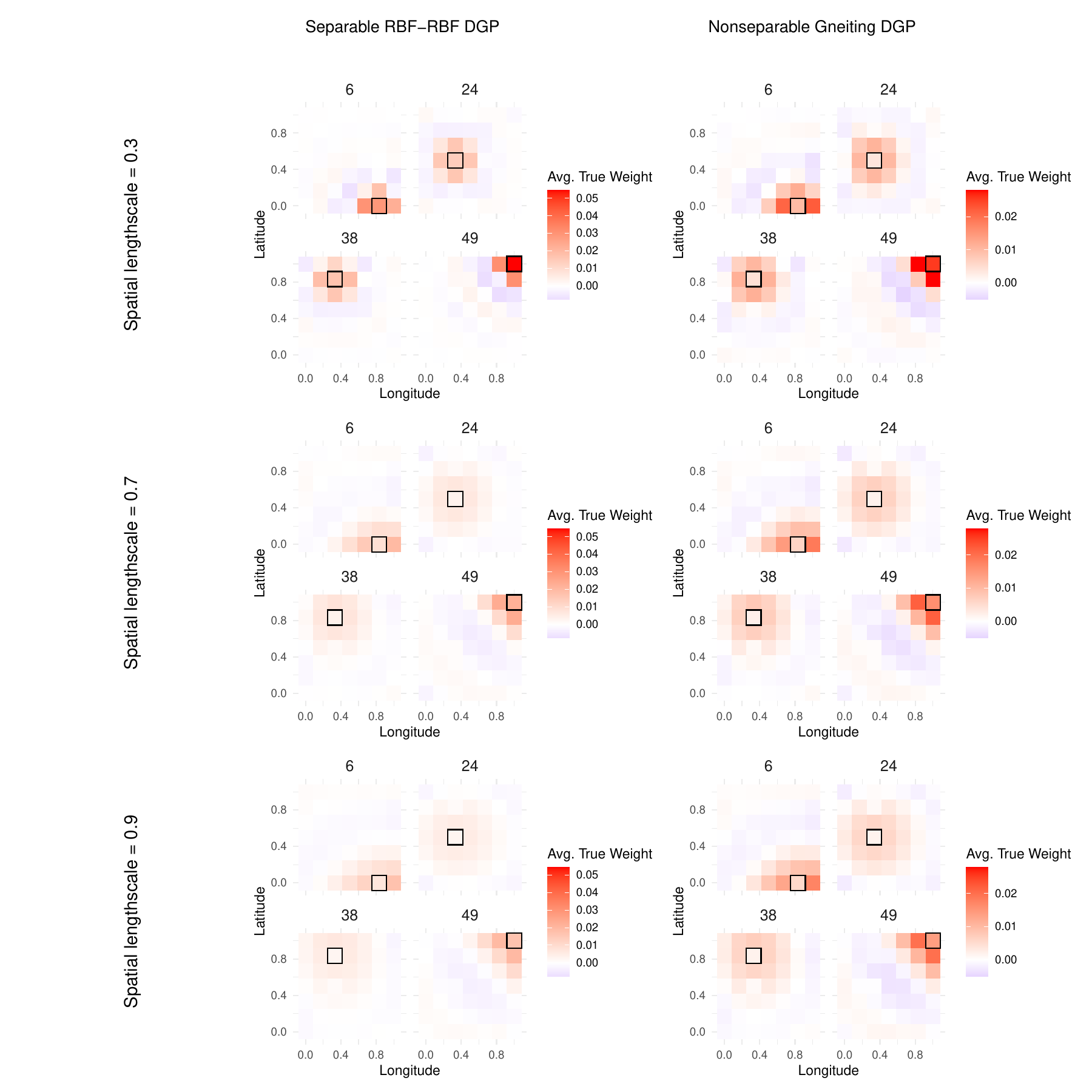}
    \caption{Average true donor weights from simulated data based on separable and nonseparable spatio-temporal kernel across spatial lengthscales. Each grid shows the spatial distribution of average true donor weights (averaged over time) for a given treated unit's counterfactual predictions. Within each of the grids, each cell corresponds to a spatial unit in the data. The cell with the black border is the treated unit in the given plot. Columns compare a separable spatio-temporal kernel (left) and a nonseparable spatio-temporal kernel (right). Rows correspond to different spatial lengthscales used during simulation: 0.3, 0.7, and 0.9. Shorter spatial lengthscales (top rows) induce more localized weight distributions, whereas longer spatial lengthscales (bottom rows) smooth weights more broadly across the spatial grid.}
    \label{fig:change_lengthscales}
\end{figure}

\begin{figure}[H]
    \centering
    \includegraphics[width=0.75\linewidth]{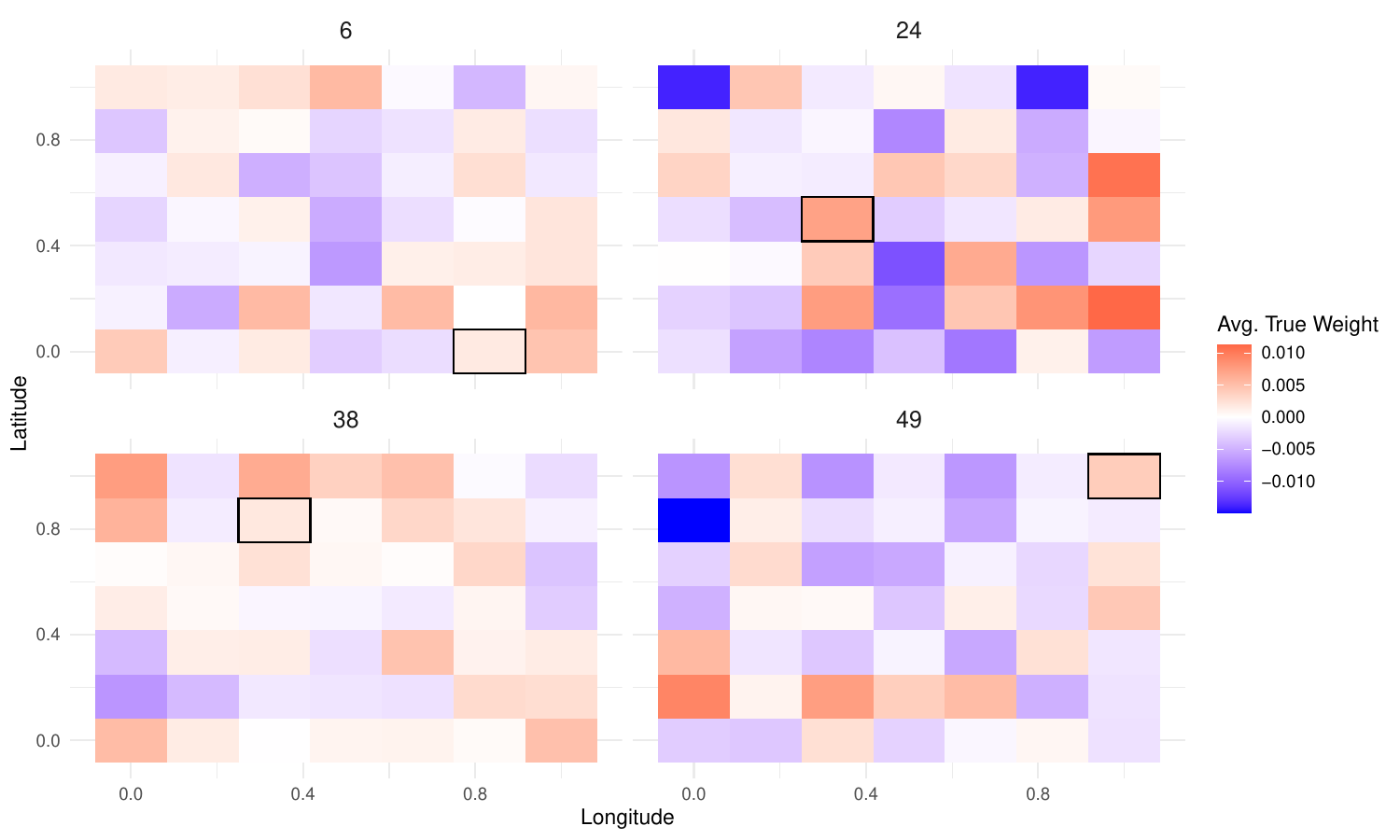}
    \caption{Average true donor weights from simulated data based on a separable ICM-RBF kernel. Each panel shows the spatial distribution of average true donor weights (averaged over time) for a given treated unit's counterfactual predictions. The cell with the black border is the treated unit in the given plot.}
    \label{fig:ICM_spatial_w}
\end{figure}

Figures \ref{fig:change_lengthscales} and \ref{fig:ICM_spatial_w} illustrate spatial distributions of donor weights averaged over time under different spatial lengthscales for separable and nonseparable kernels, highlighting how variations in kernel specifications lead to distinctly different weighting behaviors in the spatial domain. As the ICM-RBF kernel does not account for geographic proximity, we see no spatial structure in the ICM donor weights in Figure \ref{fig:ICM_spatial_w}. The separable RBF-RBF kernel and nonseparable Gneiting kernel produces spatially smooth expanding weight patterns centered around each treated unit. As the spatial lengthscale increases, the weights become more diffuse and widespread. 

\begin{figure}[H]
    \centering
    \includegraphics[width=.75\linewidth]{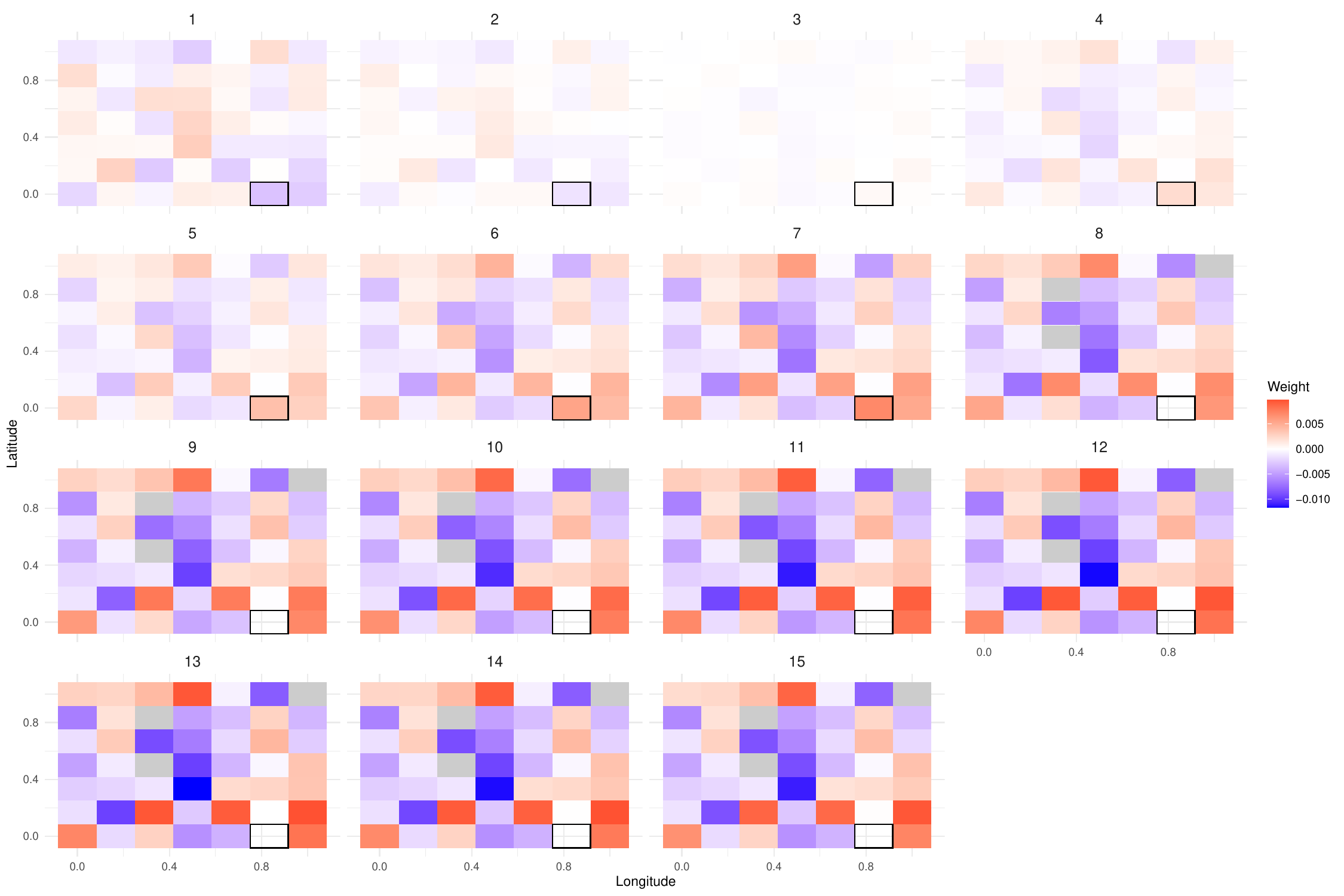}
    \caption{Spatial distribution of donor weights for Unit 6 at Time 11 using the ICM-RBF kernel. Panels show weight patterns from different time points, with red representing positive and blue representing negative weights. Grey blocks represent treated units post-treatment with zero weight.
}
    \label{fig:unit6_time_specific_ICM}
\end{figure}

\begin{figure}[H]
    \centering
    \includegraphics[width=.75\linewidth]{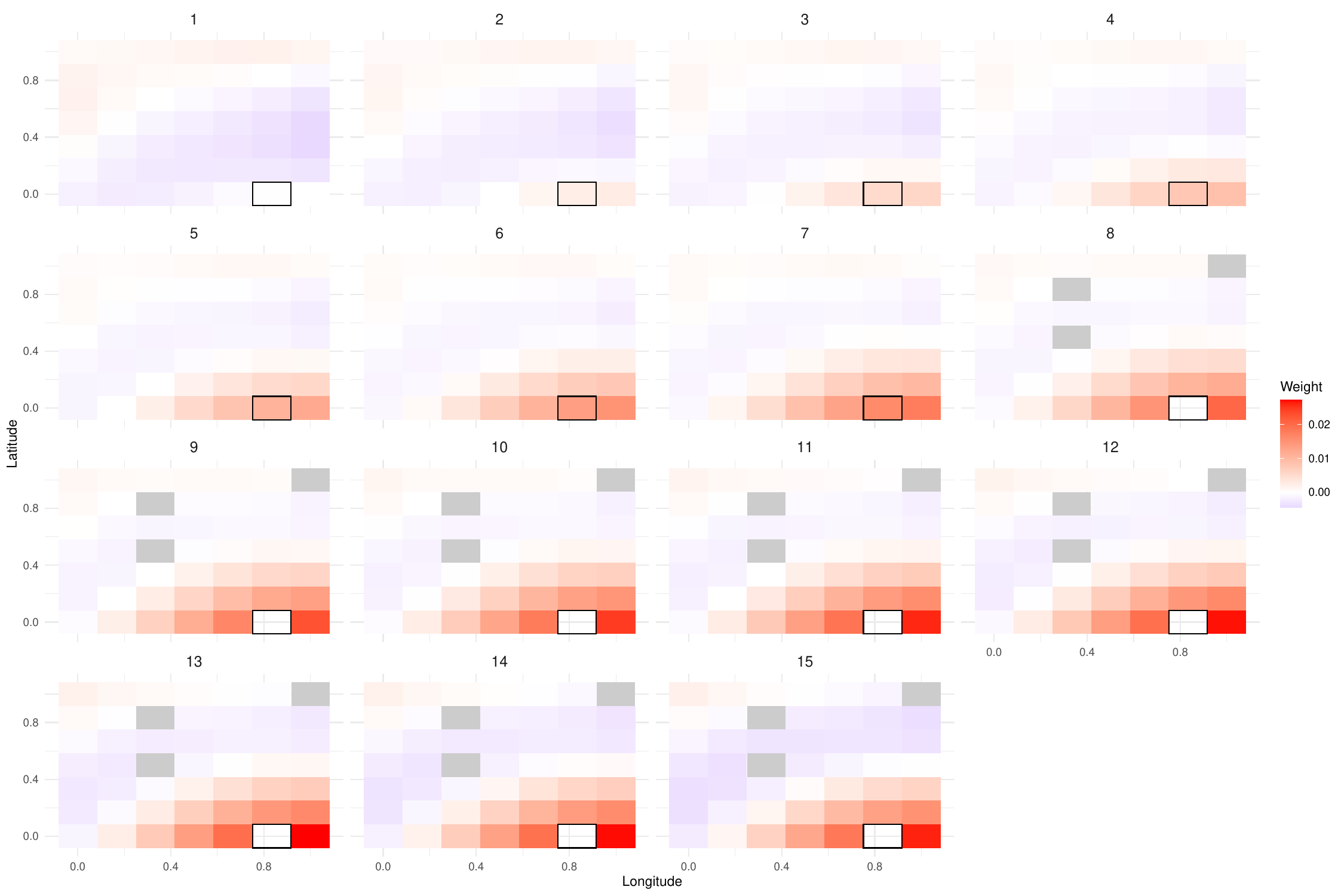}
    \caption{Spatial distribution of donor weights for Unit 6 at Time 11 using the Separable RBF-RBF kernel. Panels show weight patterns from different time points, with red representing positive and blue representing negative weights. Grey blocks represent treated units post-treatment with zero weight.}
    \label{fig:unit6_time_specific_sep}
\end{figure}

\begin{figure}[H]
    \centering
    \includegraphics[width=.75\linewidth]{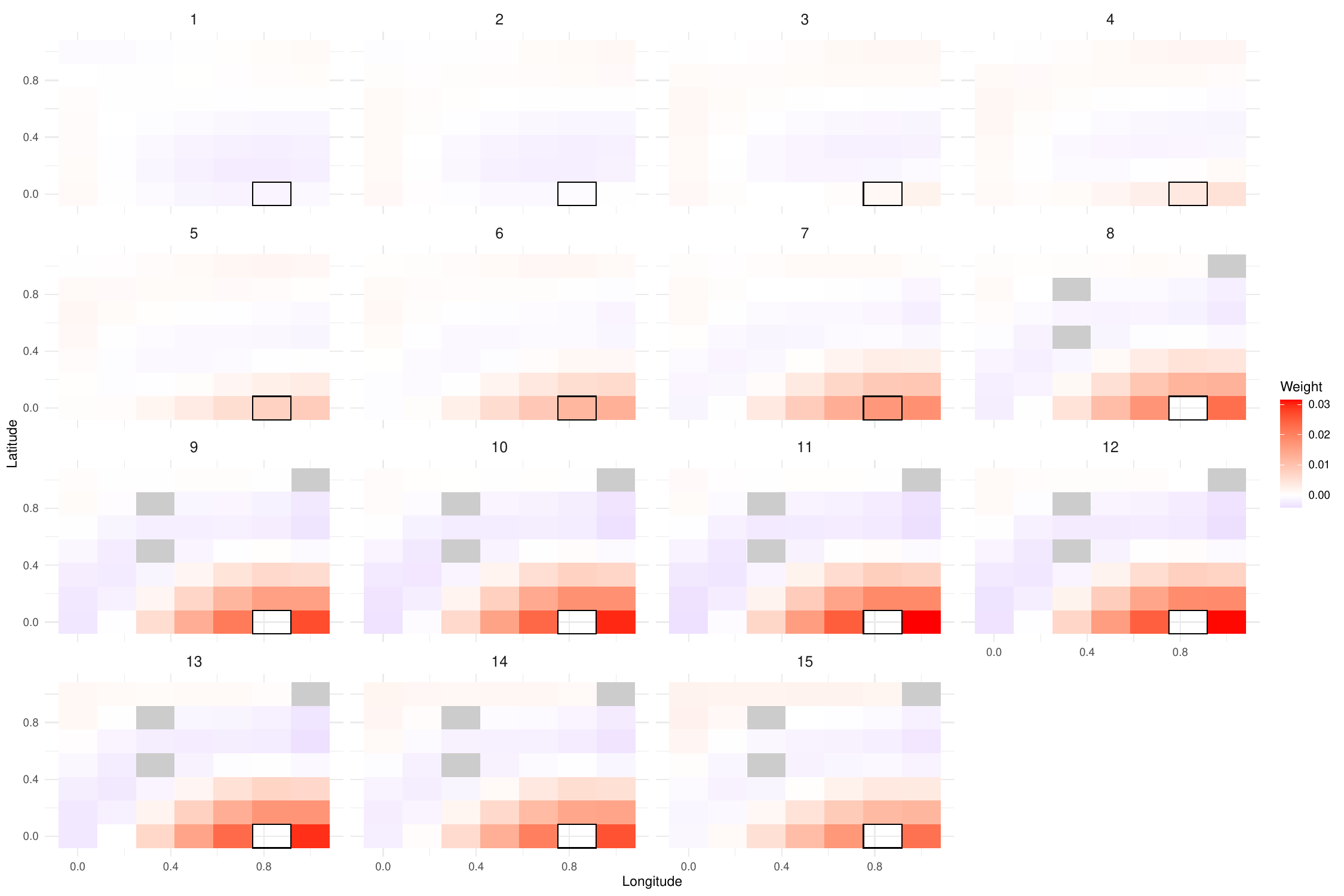}
    \caption{Spatial distribution of donor weights for Unit 6 at Time 11 using the Gneiting nonseparable kernel. Panels show weight patterns from different time points, with red representing positive and blue representing negative weights. Grey blocks represent treated units post-treatment with zero weight.
}
    \label{fig:unit6_time_specific_nonsep}
\end{figure}

To explore further the patterns of the weights over time, we focus on treated unit 6 at time 11 and plot the weights over time (Figures \ref{fig:unit6_time_specific_ICM} - \ref{fig:unit6_time_specific_nonsep}). For the separable RBF-RBF and nonseparable kernel data generating processes, we focus on the spatial lengthscale equaling 0.9. All three kernels show a decline in the magnitude of the weights as the temporal distance gets farther away from time 11, especially for the nonseparable Gneiting DGP. 

\subsection{Kernel Selection Diagnostics}

In practice, substantive knowledge about the structures of unmeasured confounders is often limited, complicating selection of an appropriate kernel in real applications of the GP approach. In particular, users may not know if the confounders in their settings have separable or non-separable spatio-temporal structures. Thus, an empirical criterion for separability may be needed to determine the most appropriate kernel selection. Several tests and diagnostics for spatio-temporal separability have been proposed in GP settings; however to our knowledge they have rarely been used in practice \cite{huang_visualization_2017, cappello_covatest_2020, allard_fully_2022, chen_space-time_2021}. One such approach is the functional boxplot approach of Huang and Sun \cite{huang_visualization_2017}, which we describe and apply in our simulated data below.

Consider a function $f_h(u)$ defined for spatial lag $h = ||s_i-s_{i'}||$ and temporal lag $u = |t-t'|$ as
$$
f_h(u)
\;=\;
\frac{k(h,u)}{k(h,0)}
\;-\;
\frac{k(0,u)}{k(0,0)},
$$
where $k(h,u)$ is the space-time covariance. Under strict separability, $f_h(u)\equiv0$ for all $h,u$. 

\begin{figure}[H]
     {\includegraphics[width=\textwidth]{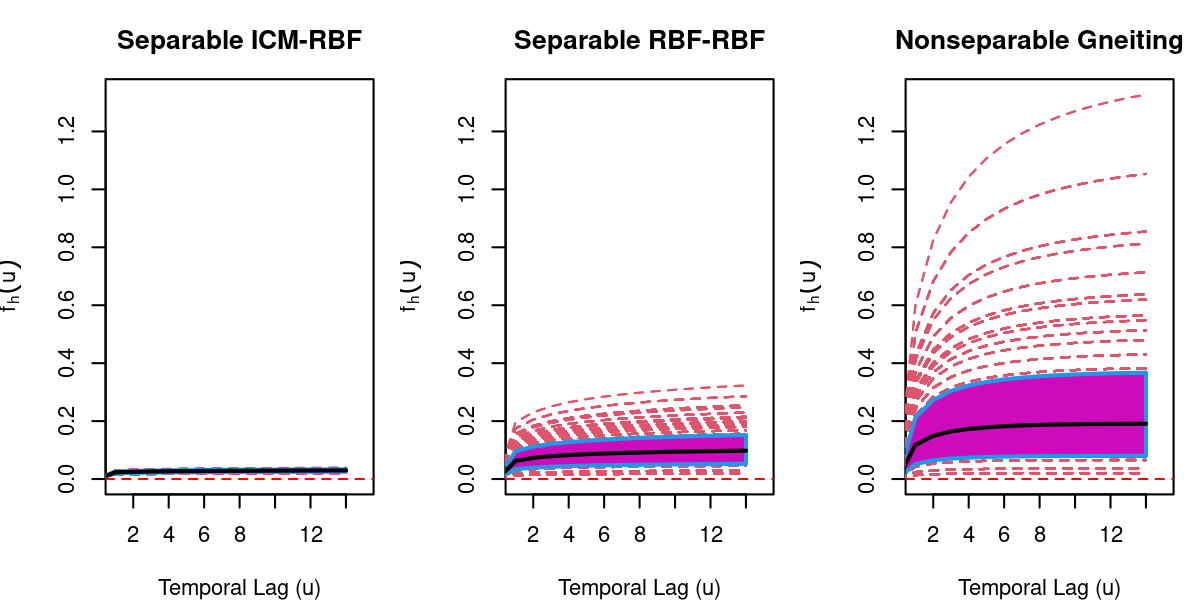}}
    \caption{Estimated functional boxplots of the separability test function for three kernel DGPs. In these plots, the functional median is shown as a central line, the 50 \% central region as a shaded band, and any extreme curves as outliers.}
    \label{fig:funcbox}
\end{figure}

In practice, $f_h(u)$ will need to be estimated from data in order to construct these plots. To emulate this scenario, we fit the nonseparable GP model to each of the simulated datasets and then use the covariance parameter estimates to estimate $f_h(u)$ and assess the functional boxplots for evidence of non-separability. In particular, after fitting the models, we extract the posterior median of the parameters in the nonseparable kernel, including the non-separability parameter $\eta$, and compute
$$
\hat f_h(u)
\;=\;
\frac{k_{\hat{\eta}}(h,u)}{k_{\hat{\eta}}(h,0)}
\;-\;
\frac{k_{\hat{\eta}}(0,u)}{k_{\hat{\eta}}(0,0)},
$$
where $k_{\eta}$ is the Gneiting covariance (Equation 3) evaluated at the estimated parameters. Because $\eta$ governs the degree of space-time interaction, its posterior concentration near zero signals separability and yields $\hat f_h(u)\approx 0$. The estimated functional boxplots are visualized in Figure \ref{fig:funcbox}. In these plots, the functional median is shown as a solid black line, the central 50\% region as a shaded magenta band, and any red extreme curves as outliers. One could also fit the nonseparable GP model and look at the proportion of MCMC samples with $\eta=0$.

Here, sampling noise and parameter uncertainty widen the central bands in all cases. The separable ICM-RBF and RBF-RBF curves show slight variability above zero reflecting a small nonzero $\hat{\eta}$. The nonseparable Gneiting estimates similarly show stronger departures from zero but with greater spread.

Other options for performing hypothesis tests of separability may exist in some contexts. For example, the \texttt{covatest} R package implements rank-based tests for separability \cite{cappello_covatest_2020}, but it requires at least 29 time points to attain adequate power. This is far more than is often available in quasi-experimental panels so these tests are likely to be under-powered. Functional boxplots may therefore provide a more practical and intuitive diagnostic for detecting departures from separability.

Another way to choose among kernels is to evaluate how well each candidate GP model captures the pre-treatment trends in the treated units. With a well-specified kernel, the model's pre-treatment predicted $Y_{it}(0)$ values should yield small prediction errors and narrow credible bands that cover the observed $Y_{it}(0)$ values, indicating that the patterns in the observed are being captured accurately. In practice, one can plot and visually assess the observed versus model-predicted $Y_{it}(0)$ or ATT values for treated units prior to treatment and/or can compute basic pre-treatment prediction error metrics. Ranking kernels' adequacy based on these pre-treatment fit statistics provides an alternative or a complement to functional boxplots and separability tests. 

\section{Simulations}\label{sec::gpsim}
To assess the performance of different GP kernel structures for estimating causal effects in spatio-temporal quasi-experiments, we conducted a series of simulations under three data-generating processes (DGPs). Each DGP corresponds to a different kernel: a nonseparable spatio-temporal Gneiting kernel, a separable spatio-temporal RBF-RBF kernel, and a separable ICM-RBF kernel. We compare counterfactual prediction performance under correctly-specified and mis-specified kernels.

Each simulated dataset consists of 49 spatial units (arranged on a $7 \times 7$ grid) observed over 15 time periods. Treatment is assigned to 10 units beginning at time point 8, and the rest remain untreated. We conduct simulations using two types of outcomes: Normally distributed outcomes and Poisson distributed outcomes. We consider a Poisson setting since health data are often reported as counts/rates, as in our motivating example (Section~\ref{sec:data}). 

For \textbf{Normal outcomes}, the data follow:
\begin{align*}
Y_{it} \sim \mathcal{N}(\mu_{it}, \sigma^2), \quad \mu_{it} = \mu_0 + f_{it}
\end{align*}

For \textbf{Poisson outcomes}, the data are generated as:
\begin{align*}
Y_{it} \sim \mathrm{Poisson}(\mu_{it}), \quad \mu_{it} = \exp(\mu_0 + f_{it})
\end{align*}
where in both cases $f_{it}$ is a Gaussian process as given in Equation~\ref{eq:gpmain}.

When generating data under the separable ICM-RBF kernel, we use a variance $\tau^2 = 0.40$, a temporal length scale $l_t = 0.90$, a low-rank spatial structure with rank $J = 5$, and an intercept $\mu_0 = 4.00$. The separable RBF-RBF kernel uses $\tau^2 = 1.00$, spatial length scale $l_s = 0.30$, temporal length scale $l_t = 0.90$, and the same intercept $\mu_0 = 4.00$. The nonseparable Gneiting kernel is parameterized with $\tau^2 = 1.00$, spatial length scale $l_s = 0.125$, temporal length scale $l_t = 0.57$, and a lower intercept $\mu_0 = 4.00$.

We generate 100 datasets for each DGP. We then fit three GP models to each simulated dataset-- a model using each of the three different kernel structures in the model fitting (one representing the correct specification and the other two a mis-specification). The outcome distribution is always correctly specified in each fitted model.

\begin{figure}[H]
    \vspace{-.5cm}\begin{center}
        \includegraphics[width=.8 \textwidth]{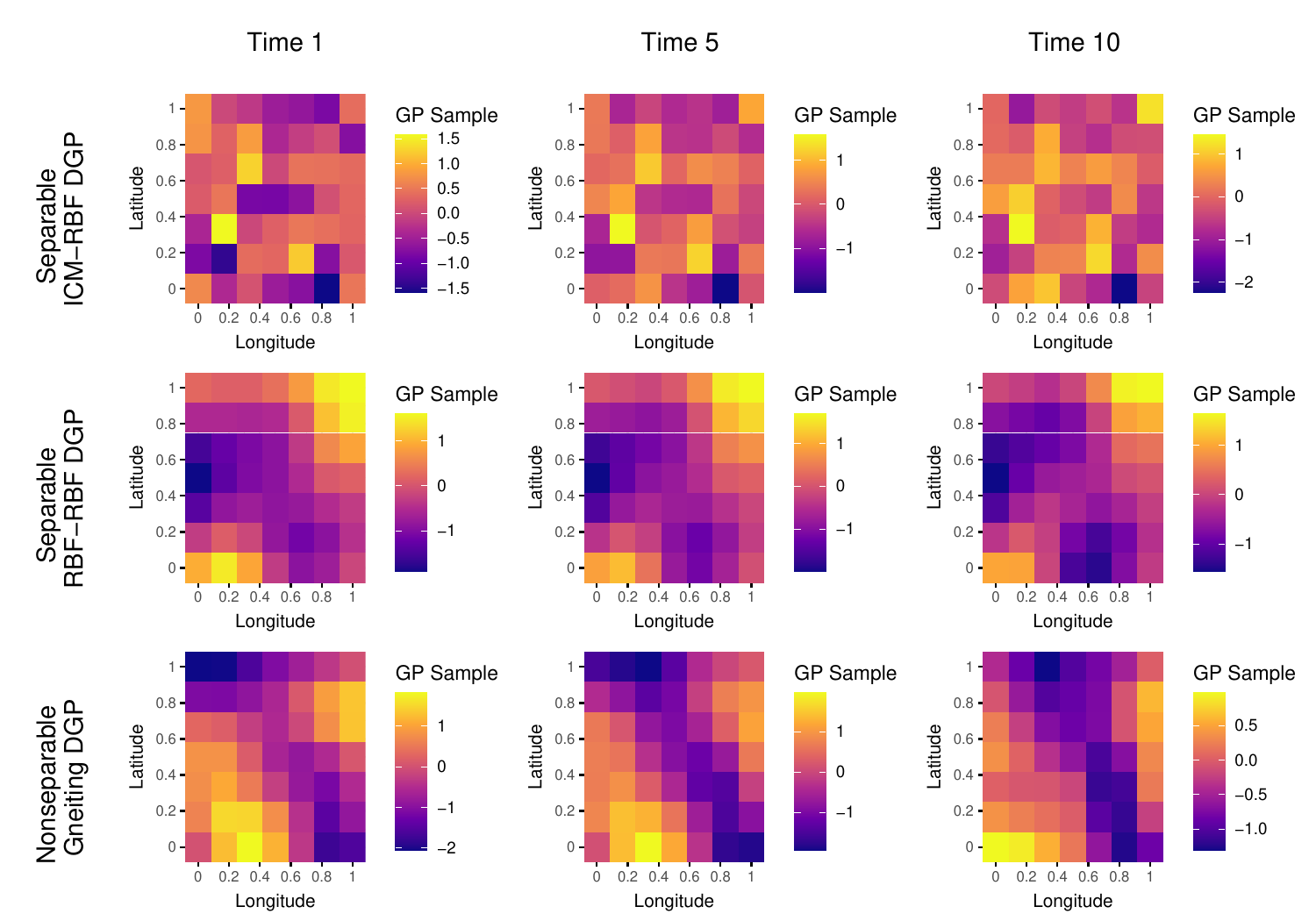}
        
    \end{center}
   \caption{Spatial domain defined over two coordinates, longitude (horizontal) and latitude (vertical), each ranging from 0 to 1. Each tile represents one draw from a Gaussian process prior at that location and time. Figure shows results for each kernel at selected time points.}
   \label{fig:gp_samples}
\end{figure}

Figure~\ref{fig:gp_samples} displays draws from the GP across selected time points. The nonseparable kernel displays complex patterns that vary dynamically over space and time, consistent with its ability to model space-time interaction. The separable RBF-RBF kernel generates smoother spatial fields with a relatively stable temporal structure over time. Alternatively, the ICM kernel produces more irregular and less interpretable spatial surfaces due to its low-rank latent structure, but displays consistency over time.

\begin{table}[!h]

\caption{Summary statistics for absolute percent bias of the counterfactuals, mean squared error (MSE), and 95\% credible interval coverage across different true kernel and model-specified kernel combinations for Normally distributed simulated data. Metrics are averaged over simulated datasets.}
\label{tab:normal_results_mean}
\centering
\begin{tabular}[t]{llrrr}
\toprule
True Kernel & Modeled Kernel & Percent Bias & MSE & Coverage\\
\midrule
\multirow{3}{10em}{Separable ICM-RBF}
 & Separable ICM-RBF & 13.38 & 0.58 & 1.00\\
& Separable RBF-RBF & 12.18 & 0.68 & 0.98\\
 & Nonseparable Gneiting & 10.73 & 0.36 & 1.00\\
\midrule
\multirow{3}{10em}{Separable RBF-RBF} & Separable ICM-RBF & 14.13 & 1.17 & 0.96\\
& Separable RBF-RBF & 9.05 & 0.31 & 1.00\\
& Nonseparable Gneiting & 5.24 & 0.09 & 1.00\\
\midrule
\multirow{3}{10em}{Nonseparable Gneiting} & Separable ICM-RBF & 18.96 & 0.70 & 1.00\\
& Separable RBF-RBF & 10.10 & 0.28 & 1.00\\
& Nonseparable Gneiting & 7.25 & 0.13 & 1.00\\
\bottomrule
\end{tabular}
\end{table}

\begin{table}[!h]

\caption{Summary statistics for absolute percent bias of the counterfactuals, mean squared error (MSE), and 95\% credible interval coverage across different true kernel and model-specified kernel combinations for Poisson distributed simulated data. Metrics are averaged over simulated datasets.}
\label{tab:poisson_results_mean}
\centering
\begin{tabular}[t]{llrrr}
\toprule
True Kernel & Modeled Kernel & Percent Bias & MSE & Coverage\\
\midrule
\multirow{3}{10em}{Separable ICM-RBF} & Separable ICM-RBF & 10.44 & 311.23 & 1.00\\
& Separable RBF-RBF & 47.36 & 5101.96 & 0.31\\
 & Nonseparable Gneiting & 26.80 & 1267.09 & 0.90\\
\midrule
\multirow{3}{10em}{Separable RBF-RBF}& Separable ICM-RBF & 14.34 & 2122.98 & 0.98\\
& Separable RBF-RBF & 22.17 & 1858.03 & 0.59\\
& Nonseparable Gneiting & 9.56 & 494.06 & 0.97\\
\midrule
\multirow{3}{10em}{Nonseparable Gneiting} & Separable ICM-RBF  & 28.21 & 474.27 & 0.98\\
& Separable RBF-RBF & 26.09 & 271.60 & 0.59\\
& Nonseparable Gneiting & 13.14 & 71.51 & 1.00\\
\bottomrule
\end{tabular}
\end{table}

Performance across models is summarized in Tables~\ref{tab:normal_results_mean} (Normal) and~\ref{tab:poisson_results_mean} (Poisson), reporting absolute percent bias of the counterfactuals, mean squared error (MSE), and 95\% credible interval coverage. In the Normal distribution simulations, regardless of the true kernel used in the DGP, bias and MSE are the lowest when the Nonseparable kernel is used in model-fitting, followed by the separable RBF-RBF model. These results suggest that for Normal outcomes, specifying the non-separable kernel in model fitting maybe a ``safe'' choice regardless of the true structure of the unmeasured confounders. While the ICM-RBF kernel always had the highest bias, its performance notably deteriorated relative to the other kernels when the true DGP exhibited spatial smoothness. Under the nonseparable Gneiting DGP, its mean bias reached 18.96 and MSE rose to 0.70. These results indicate that the ICM's low-rank latent factorization may fail to capture spatially structured variation adequately, leading to poor counterfactual estimates. Credible interval coverage was consistently conservative across settings.

For Poisson outcomes, bias and MSEs are generally higher than in the Normal simulations, and credible interval coverage is generally lower. When the true kernel is the separable ICM-RBF or the non-separable, the best performance is achieved when the model-specified kernel matches the true kernel. When the true kernel is the separable RBF-RBF, the models fit using the non-separable kernel perform best. Interestingly, the models using the nonseparable kernel perform robustly across all scenarios, with moderate bias and strong coverage even under misspecification. When there is spatial structure in the data, the nonseparable kernel model always performs best. This suggests the nonseparable kernel is flexible in capturing diverse spatio-temporal structures in the Normal and Poisson outcome settings.

\section{Application}
We applied the GP approach with each of the three kernels to the Hurricane Katrina exposure and Medicare mortality data. Because the outcomes are county-level counts of mortality, we use a Poisson likelihood and a log link, i.e., 
$$
        \mathrm{log}(\mathrm{E}[Y_{it}(0)]) = \mu_0 + \delta_{i} + f_{it} +\mathrm{log}(\theta_{it}), \quad \mathbf{f} \sim \mathcal{G}\mathcal{P}(0, \mathbf{K})
$$
where $Y_{it}(0)$ are the mortality counts under control, $\mu_0$ is a global intercept, $\delta_i$ is a county-specific intercept (fixed effect), and $\theta_{it}$ is the Medicare population for county $i$ and time $t$. Including $\delta_i$ helps control for differences in baseline mortality rates across counties that may not be fully captured by the smooth GP term alone. We fit models specifying each of the separable ICM-RBF kernel, the separable RBF-RBF kernel, and the nonseparable Gneiting kernel. For each model, we ran 1,000 Hamiltonian Monte Carlo iterations in \texttt{rstan}, discarding the first 500 as burn-in. We report posterior estimates of the ATT, averaged across all treated counties.

Table \ref{tab:rhat} reports the average R-hat value, a commonly-used measure of Bayesian model convergence, for the counterfactual predictions of each model. Values less than 1.05 indicate that the corresponding posterior chains have likely converged \cite{vehtari_rank-normalization_2021}. All models displayed average R-hat values below this threshold, suggesting adequate convergence to the posterior predictive distribution of the counterfactuals. Similarly, Figure \ref{fig:trace_plots} displays trace plots of the posterior mean of the predicted counts averaged across all treated counties at time point $t=10$, for each GP model. The plots agree with the R-hat values in that the Markov chains are well-mixed.

\begin{table}[H]
\centering
\caption{
Posterior median estimates of the overall Average Treatment Effect on the Treated (ATT), with 95\% credible intervals, for each Gaussian Process model. ATT values represent the average difference between observed and model-estimated counterfactual mortality rates per 100,000 across all treated counties during a two week period surrounding Hurricane Katrina. 
}
\label{tab:att_summary}

\begin{tabular}{lcc}
  \hline
Model & ATT & 95\% CI \\ 
  \hline
Separable RBF-ICM & 83.0 & (63.1, 99.6) \\ 
Separable RBF-RBF & 82.4 & (62.8, 99.9) \\ 
Nonseparable Gneiting & 79.7 & (62.4, 99.2) \\ 
   \hline
\end{tabular}

\end{table}

\begin{figure}[H]
  \centering
    \includegraphics[width=\linewidth]{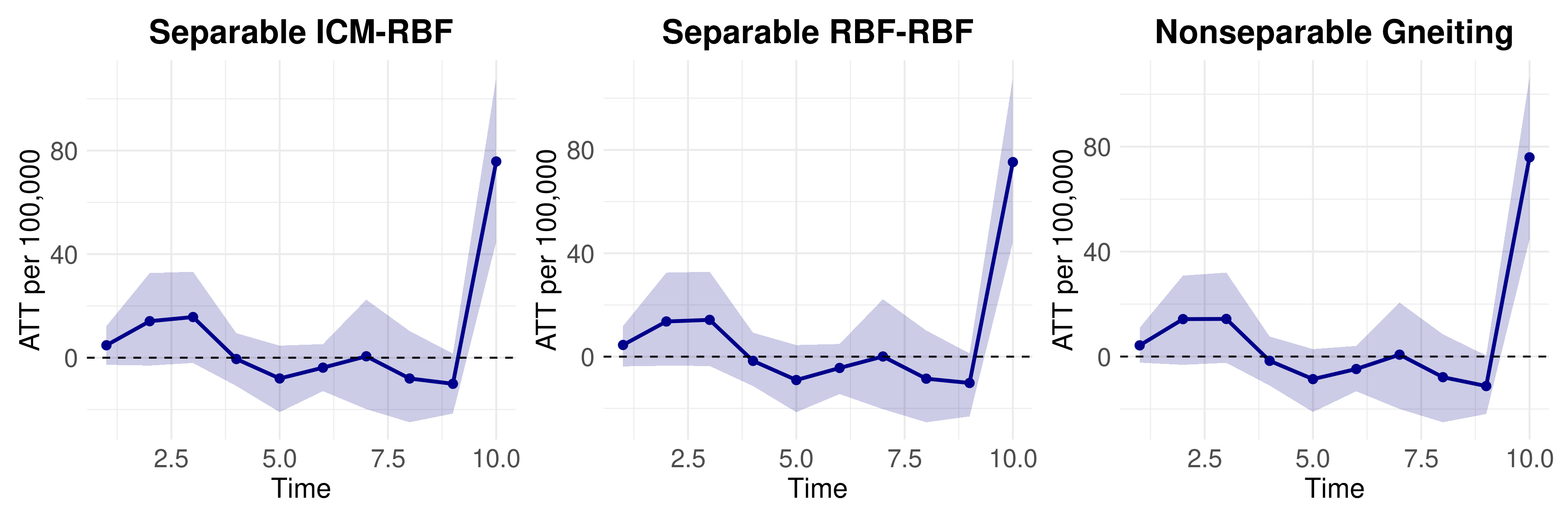}
  \caption{
Posterior median estimates of the Average Treatment Effect on the Treated (ATT) over time, with 95\% credible intervals for each model. 
ATT values are computed as the difference between observed mortality rates and model-estimated mortality rates under control (per 100,000) for treated counties at each time point. Shaded bands represent time point-specific 95\% credible intervals.
}
  \label{fig:att}
\end{figure}

Figure \ref{fig:funcbox_katrina} displays the estimated functional boxplot for the Katrina mortality data. Such minimal deviation indicates that the data exhibit a nearly separable space-time covariance structure, implying that a separable kernel is sufficient here. Figure \ref{fig:att} shows the time period-specific ATT estimates,
$$\widehat{ATT}_t=\frac{1}{N_1} \sum_{i: D_{i} = 1} \left[ Y_{it} - \widehat{Y}_{it}(0) \right]$$
for each model. Here, we can get an idea of how well the models capture the pre-treatment trends in the treated counties (generally considered an indicator of good model fit). Prior to treatment (Time 10) the ATT estimates are generally centered around zero suggesting adequate model fit for all models, further supporting the adequacy of a separable specification in this application. Time point 10, the two-week period immediately following Katrina's landfall, shows a sharp increase in the ATT for all models with credible intervals separating from zero. This indicates an increase in mortality post-storm.

The overall estimated ATTs and 95\% credible intervals are shown in Table \ref{tab:att_summary}. All models show a substantial increase in mortality attributable to the hurricane. The separable ICM-RBF model estimates an average increase of 83.0 deaths per 100,000 (95
\% CI: 63.1, 99.6) and the separable RBF-RBF model estimates an average increase of 82.4 deaths per 100,000 (95\% CI: 62.8, 99.9) due to the storm. The nonseparable Gneiting model similarly estimates an average increase of 79.7 deaths per 100,000 (95\% CI: 62.4, 99.2) due to the storm. The 95\% credible intervals exclude zero, indicating strong evidence of a causal effect of Hurricane Katrina on mortality among Medicare enrollees. These results provide evidence that Hurricane Katrina led to a statistically significant increase in short-term mortality among Medicare beneficiaries in the affected counties, agreeing with many previous studies \cite{stephens2007excess,brunkard_hurricane_2008,parks2023short,nethery_integrated_2023}. 

\section{Discussion}

In this paper, we demonstrated that GPs provide a flexible framework for conducting causal inference in quasi-experimental settings with spatio-temporal data. By encoding assumptions about unmeasured confounding directly into the covariance structure, GPs enable the estimation of counterfactual outcomes while accounting for complex dependencies across units and time. We highlighted how the GP framework unifies and extends existing approaches, offering both interpretability and modeling flexibility. Our simulation study highlighted that kernel choice has considerable implications for bias. When the specified kernel matched the data generating process, we observed minimal absolute percent bias of the counterfactuals. Importantly, models using the nonseparable Gneiting kernel performed robustly across all scenarios, even under misspecification.

We also provided guidance for choosing between separable and nonseparable kernels and illustrated how different kernels impose different smoothing structures through simulations and weighting illustrations. These insights make the GP approach more interpretable and easier to tailor to specific applications. All code used in this paper is publicly available \url{https://github.com/sofiavega98/spacetimegp/}. Future work should explore computational strategies, such as nearest neighbor approximations to scale GPs to larger datasets without sacrificing flexibility or interpretability.

\section*{Acknowledgements}

Support for this research was provided by NIH grant K01ES032458, the Harvard Climate Change Solutions Fund, and the Harvard Milton Fund. The computations in this paper were performed on the FASRC Cannon cluster and the FAS Secure Environment cluster, supported by the Faculty of Arts and Sciences Division of Science Research Computing Group at Harvard University.

\section*{Data and code availability}

Medicare enrollee data can be publicly accessed after purchasing and completing an application process through CMS. Hurricane exposure data is publicly available through the \texttt{hurricaneexposuredata} R package \cite{ anderson_assessing_2020}. Code to reproduce results is available at \\ \url{https://github.com/sofiavega98/spacetimegp/}.

\newpage
\bibliographystyle{unsrt}
\bibliography{bib}

\begin{thebibliography}{10}

\bibitem{li2021multi}
Didong Li, Andrew Jones, Sudipto Banerjee, and Barbara~E Engelhardt.
\newblock Multi-group gaussian processes.
\newblock {\em arXiv preprint arXiv:2110.08411}, 2021.

\bibitem{ren2021bayesian}
Boyu Ren, Xiao Wu, Danielle Braun, Natesh Pillai, and Francesca Dominici.
\newblock A bayesian gaussian process for estimating a causal exposure response
  curve in environmental epidemiology.
\newblock {\em arXiv preprint arXiv:2105.03454}, 2021.

\bibitem{ben2023estimating}
Eli Ben-Michael, David Arbour, Avi Feller, Alexander Franks, and Steven
  Raphael.
\newblock Estimating the effects of a california gun control program with
  multitask gaussian processes.
\newblock {\em The Annals of Applied Statistics}, 17(2):985--1016, 2023.

\bibitem{brunkard_hurricane_2008}
Joan Brunkard, Gonza Namulanda, and Raoult Ratard.
\newblock Hurricane {Katrina} {Deaths}, {Louisiana}, 2005.
\newblock {\em Disaster Medicine and Public Health Preparedness},
  2(4):215--223, December 2008.

\bibitem{nethery_integrated_2023}
Rachel~C Nethery, Nina Katz-Christy, Marianthi-Anna Kioumourtzoglou, Robbie~M
  Parks, Andrea Schumacher, and G~Brooke Anderson.
\newblock Integrated causal-predictive machine learning models for tropical
  cyclone epidemiology.
\newblock {\em Biostatistics}, 24(2):449--464, April 2023.

\bibitem{anderson_assessing_2020}
G.~Brooke Anderson, Joshua Ferreri, Mohammad Al-Hamdan, William Crosson, Andrea
  Schumacher, Seth Guikema, Steven Quiring, Dirk Eddelbuettel, Meilin Yan, and
  Roger~D. Peng.
\newblock Assessing {United} {States} {County}-{Level} {Exposure} for
  {Research} on {Tropical} {Cyclones} and {Human} {Health}.
\newblock {\em Environmental Health Perspectives}, 128(10):107009, October
  2020.

\bibitem{yan_tropical_2021}
Meilin Yan, Ander Wilson, Francesca Dominici, Yun Wang, Mohammad Al-Hamdan,
  William Crosson, Andrea Schumacher, Seth Guikema, Sheryl Magzamen,
  Jennifer~L. Peel, Roger~D. Peng, and G.~Brooke Anderson.
\newblock Tropical {Cyclone} {Exposures} and {Risks} of {Emergency} {Medicare}
  {Hospital} {Admission} for {Cardiorespiratory} {Diseases} in 175 {Urban}
  {United} {States} {Counties}, 1999-2010.
\newblock {\em Epidemiology}, 32(3):315--326, May 2021.

\bibitem{Rubin1974}
Donald~B. Rubin.
\newblock {Estimating causal effects of treatments in randomized and
  nonrandomized studies.}
\newblock {\em {Journal of Educational Psychology}}, 66(5):688--701, October
  1974.

\bibitem{pang_bayesian_2022}
Xun Pang, Licheng Liu, and Yiqing Xu.
\newblock A {Bayesian} {Alternative} to {Synthetic} {Control} for {Comparative}
  {Case} {Studies}.
\newblock {\em Political Analysis}, 30(2):269--288, April 2022.

\bibitem{rubin1980randomization}
Donald~B Rubin.
\newblock Randomization analysis of experimental data: The {Fisher}
  randomization test comment.
\newblock {\em {Journal of the American Statistical Association}},
  75(371):591--593, 1980.

\bibitem{reich2021review}
Brian~J Reich, Shu Yang, Yawen Guan, Andrew~B Giffin, Matthew~J Miller, and Ana
  Rappold.
\newblock A review of spatial causal inference methods for environmental and
  epidemiological applications.
\newblock {\em International Statistical Review}, 89(3):605--634, 2021.

\bibitem{gneiting2002nonseparable}
Tilmann Gneiting.
\newblock Nonseparable, stationary covariance functions for space-time data.
\newblock {\em Journal of the American Statistical Association},
  97(458):590--600, 2002.

\bibitem{suryasentana2023demystifying}
SK~Suryasentana and BB~Sheil.
\newblock Demystifying the connections between gaussian process regression and
  kriging theories.
\newblock In {\em Offshore Site Investigation Geotechnics 9th International
  Conference Proceeding}, volume 2020, pages 2020--2027. Society for Underwater
  Technology, 2023.

\bibitem{huang_visualization_2017}
Huang Huang and Ying Sun.
\newblock Visualization and {Assessment} of {Spatio}-temporal {Covariance}
  {Properties}, May 2017.
\newblock arXiv:1705.01789 [stat].

\bibitem{cappello_covatest_2020}
Claudia Cappello, Sandra De~Iaco, and Donato Posa.
\newblock \textbf{covatest} : {An} \textit{{R}} {Package} for {Selecting} a
  {Class} of {Space}-{Time} {Covariance} {Functions}.
\newblock {\em Journal of Statistical Software}, 94(1), 2020.

\bibitem{allard_fully_2022}
Denis Allard, Lucia Clarotto, and Xavier Emery.
\newblock Fully nonseparable {Gneiting} covariance functions for multivariate
  space-time data.
\newblock {\em Spatial Statistics}, 52:100706, December 2022.

\bibitem{chen_space-time_2021}
Wanfang Chen, Marc~G. Genton, and Ying Sun.
\newblock Space-{Time} {Covariance} {Structures} and {Models}.
\newblock {\em Annual Review of Statistics and Its Application}, 8(1):191--215,
  March 2021.

\bibitem{vehtari_rank-normalization_2021}
Aki Vehtari, Andrew Gelman, Daniel Simpson, Bob Carpenter, and Paul-Christian
  Bürkner.
\newblock Rank-{Normalization}, {Folding}, and {Localization}: {An} {Improved}
  {Rhat} for {Assessing} {Convergence} of {MCMC} (with {Discussion}).
\newblock {\em Bayesian Analysis}, 16(2), June 2021.

\bibitem{stephens2007excess}
Kevin~U Stephens~Sr, David Grew, Karen Chin, Paul Kadetz, P~Gregg Greenough,
  Frederick~M Burkle~Jr, Sandra~L Robinson, and Evangeline~R Franklin.
\newblock Excess mortality in the aftermath of hurricane katrina: a preliminary
  report.
\newblock {\em Disaster medicine and public health preparedness}, 1(1):15--20,
  2007.

\bibitem{parks2023short}
Robbie~M Parks, Vasilis Kontis, G~Brooke Anderson, Jane~W Baldwin, Goodarz
  Danaei, Ralf Toumi, Francesca Dominici, Majid Ezzati, and Marianthi-Anna
  Kioumourtzoglou.
\newblock Short-term excess mortality following tropical cyclones in the united
  states.
\newblock {\em Science advances}, 9(33):eadg6633, 2023.

\end{thebibliography}

\newpage
\pagenumbering{arabic}
\renewcommand*{\thepage}{S\arabic{page}}
\beginsupplement
\section{Supplementary Materials}
\setcounter{figure}{0}
\setcounter{table}{0}

\subsection{Gaussian-Process Model Specifications}\label{sec::priors}

\subsubsection{Simulations}
Let units $i=1,\dots,N$ and times $t=1,\dots,T$.  Denote the latent process $f_{it}$ and observed outcome $Y_{it}$. We fit three different covariance kernels, each with both a Normal and a Poisson-count likelihood.

\bigskip
\noindent\textbf{GP prior:}
$$
\mathbf{f} \sim \mathcal{GP}\bigl(0,\;k\bigl((i,t),(i',t')\bigr)\bigr).
$$
\newline
\noindent\textbf{1. Nonseparable (Gneiting) kernel}
Define
$$
\psi(u) = \left(\frac{1}{l_t}\,u^\alpha + 1\right)^{\eta}, 
\quad
\phi(z) = \exp\left(-\frac{1}{t_s}\,z^\gamma\right),
$$
with variance $\tau^2>0$, length-scales $l_t,l_s>0$, smoothness $\alpha,\gamma\in(0,1]$, and space-time interaction $\eta\in[0,1]$.  Then
$$
k^{\mathrm{G}}\bigl((i,t),(i',t')\bigr)
= \tau^2\,
\psi\bigl(|t-t'|^2\bigr)\,
\phi\!\Bigl(\tfrac{\lVert s_i - s_{i'}\rVert^2}
                     {\psi\bigl(|t-t'|^2\bigr)}\Bigr).
$$

\noindent\textbf{2. Separable RBF-RBF kernel}
\[
k^{\mathrm{RBF-RBF}}\bigl((i,t),(i',t')\bigr)
= \tau^2\,
\exp\!\Bigl(-\tfrac{\lVert s_i - s_{i'}\rVert^2}{2l_s^2}\Bigr)\,
\exp\!\Bigl(-\tfrac{|t - t'|^2}{2l_t^2}\Bigr).
\]

\noindent\textbf{3. Separable ICM-RBF kernel}
Let \(\Phi\in\mathbb R^{N\times J}\) have rows \(\phi_i\).  Then
\[
k^{\mathrm{ICM-RBF}}\bigl((i,t),(i',t')\bigr)
= (\phi_i^{\!\top}\!\phi_{i'})\;
\exp\!\Bigl(-\tfrac{|t - t'|^2}{2l_t^2}\Bigr).
\]

\bigskip
\noindent\textbf{Outcome models:}
\[
\text{(a) Normal:}
\quad Y_{it}\mid f_{it}\;\sim\;\mathcal{N}\bigl(\mu_0 + f_{it},\,\sigma^2\bigr),
\qquad
\text{(b) Poisson:}
\quad Y_{it}\mid f_{it}\;\sim\;\mathrm{Poisson}\!\bigl(\exp(\mu_0 + f_{it})\bigr).
\]

\bigskip
\noindent\textbf{Priors (all models):}
$$
\tau^2,\;\sigma^2,\;l_t,\;l_s
\;\sim\;\mathrm{Inv\!-\!Gamma}(5,5),
\quad
\alpha,\;\gamma\;\sim\;\mathrm{Unif}(0,1),\\
\quad
\eta \sim \mathcal{N}(0.5,0.1)
\quad
\mu_0\;\sim\;\mathrm{Unif}\bigl(-\infty,\infty\bigr).
$$

\subsubsection{Application}
Let units $i=1,\dots,N$ and times $t=1,\dots,T$.  Denote the latent process $f_{it}$ and observed outcome $Y_{it}$. We fit three different covariance kernels on the Katrina mortality data with a Poisson-count likelihood. Here, we adjust the priors based on the county-distances and mortality counts of our data.

\bigskip
\noindent\textbf{GP prior:}
$$
\mathbf{f} \sim \mathcal{GP}\bigl(0,\;k\bigl((i,t),(i',t')\bigr)\bigr).
$$

\noindent\textbf{1. Nonseparable (Gneiting) kernel}
Define
$$
\psi(u) = \left(\frac{1}{l_t}\,u^\alpha + 1\right)^{\eta}, 
\quad
\phi(z) = \exp\left(-\frac{1}{t_s}\,z^\gamma\right),
$$
with variance $\tau^2>0$, length-scales $l_t,l_s>0$, smoothness $\alpha,\gamma\in(0,1]$, and space-time interaction $\eta\in[0,1]$.  Then
$$
k^{\mathrm{G}}\bigl((i,t),(i',t')\bigr)
= \tau^2\,
\psi\bigl(|t-t'|^2\bigr)\,
\phi\!\Bigl(\tfrac{\lVert s_i - s_{i'}\rVert^2}
                     {\psi\bigl(|t-t'|^2\bigr)}\Bigr).
$$

\noindent\textbf{2. Separable RBF-RBF kernel}
\[
k^{\mathrm{RBF-RBF}}\bigl((i,t),(i',t')\bigr)
= \tau^2\,
\exp\!\Bigl(-\tfrac{\lVert s_i - s_{i'}\rVert^2}{2l_s^2}\Bigr)\,
\exp\!\Bigl(-\tfrac{|t - t'|^2}{2l_t^2}\Bigr).
\]

\noindent\textbf{3. Separable ICM-RBF kernel}
Let \(\Phi\in\mathbb R^{N\times J}\) have rows \(\phi_i\).  Then
\[
k^{\mathrm{ICM-RBF}}\bigl((i,t),(i',t')\bigr)
= (\phi_i^{\!\top}\!\phi_{i'})\;
\exp\!\Bigl(-\tfrac{|t - t'|^2}{2l_t^2}\Bigr).
\]

\bigskip
\noindent\textbf{Outcome model:}

\[
\text{Poisson:}
\quad Y_{it}\mid f_{it}\;\sim\;\mathrm{Poisson}\!\bigl(\exp(\mu_0 + \delta_{i} + f_{it} +\mathrm{log}(\theta_{it}))\bigr).
\]

\bigskip
\noindent\textbf{Priors (Nonseparable):}
$$
\tau^2,\;\mu_0
\;\sim\;\mathcal{N}(0,1), 
\quad
\;l_t, \sim \mathcal{N}(10, 5), 
\quad
\;l_s \sim \mathcal{N}(300, 100),
\quad
\alpha,\;\gamma,\;\eta\sim\;\mathrm{Beta}(1,1),\\
\quad
\delta_i\;\sim\;\mathrm{Unif}\bigl(-\infty,\infty\bigr)
$$

\noindent\textbf{Priors (Separable RBF-RBF):}
$$
\tau^2 
\;\sim\;\mathcal{N}(0,1), 
\quad
\;l_t, \sim N(10, 5), 
\quad
\;l_s \sim N(300, 100),
\quad
\mu_0,\;\delta_i\;\sim\;\mathrm{Unif}\bigl(-\infty,\infty\bigr)
$$

\noindent\textbf{Priors (Separable ICM-RBF):}
$$
\tau^2 
\;\sim\;\mathrm{Inv\!-\!Gamma}(5,5), 
\quad
\;l_t, \sim N(10, 5), 
\quad
\;l_s \sim N(300, 100),
\quad
\mu_0,\;\sim\;\mathrm{Unif}\bigl(-\infty,\infty\bigr),
\quad
\delta_i\sim\mathcal{N}(0,1)
$$

\subsection{Additional Tables and Figures}

\begin{figure}
    \centering
    \includegraphics[width=1\linewidth]{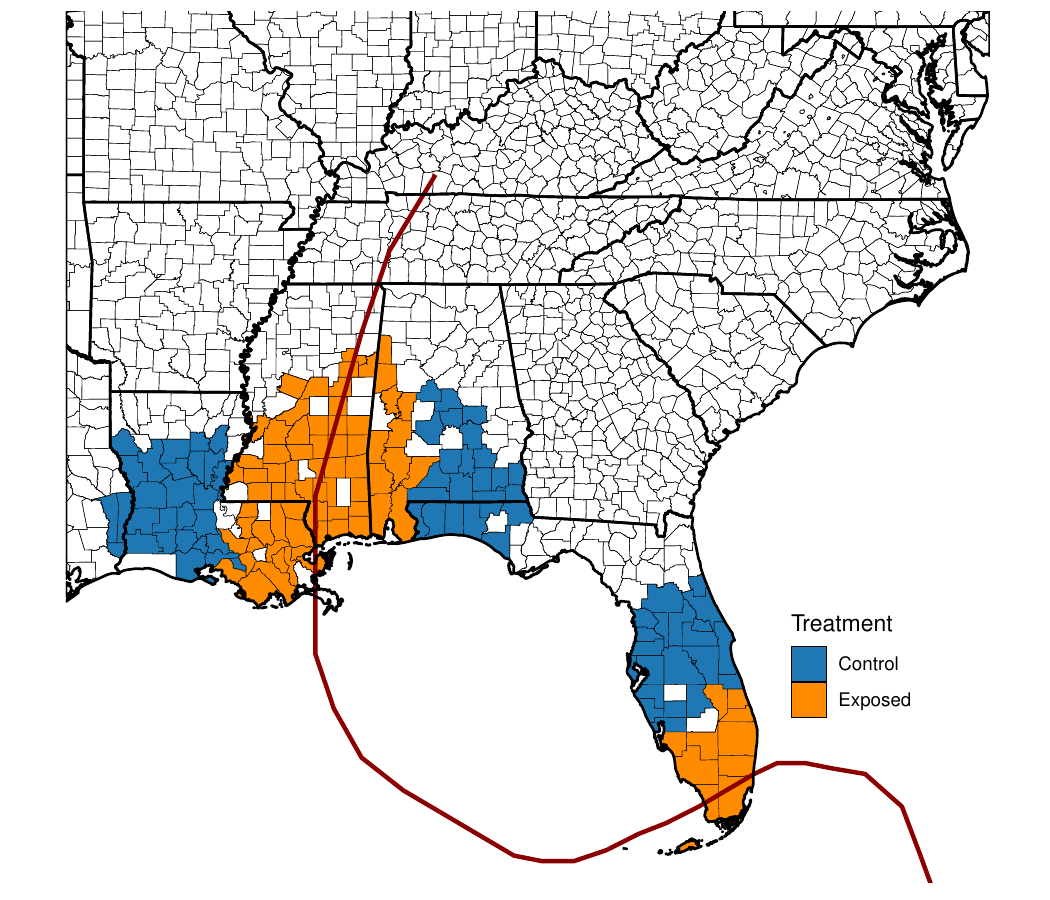}
    \caption{Map of the exposed and control counties used in the Hurricane Katrina mortality case-study. The red line indicates storm path.}
    \label{fig::map}
\end{figure}

\begin{figure}
    \centering
    \includegraphics[width=1\linewidth]{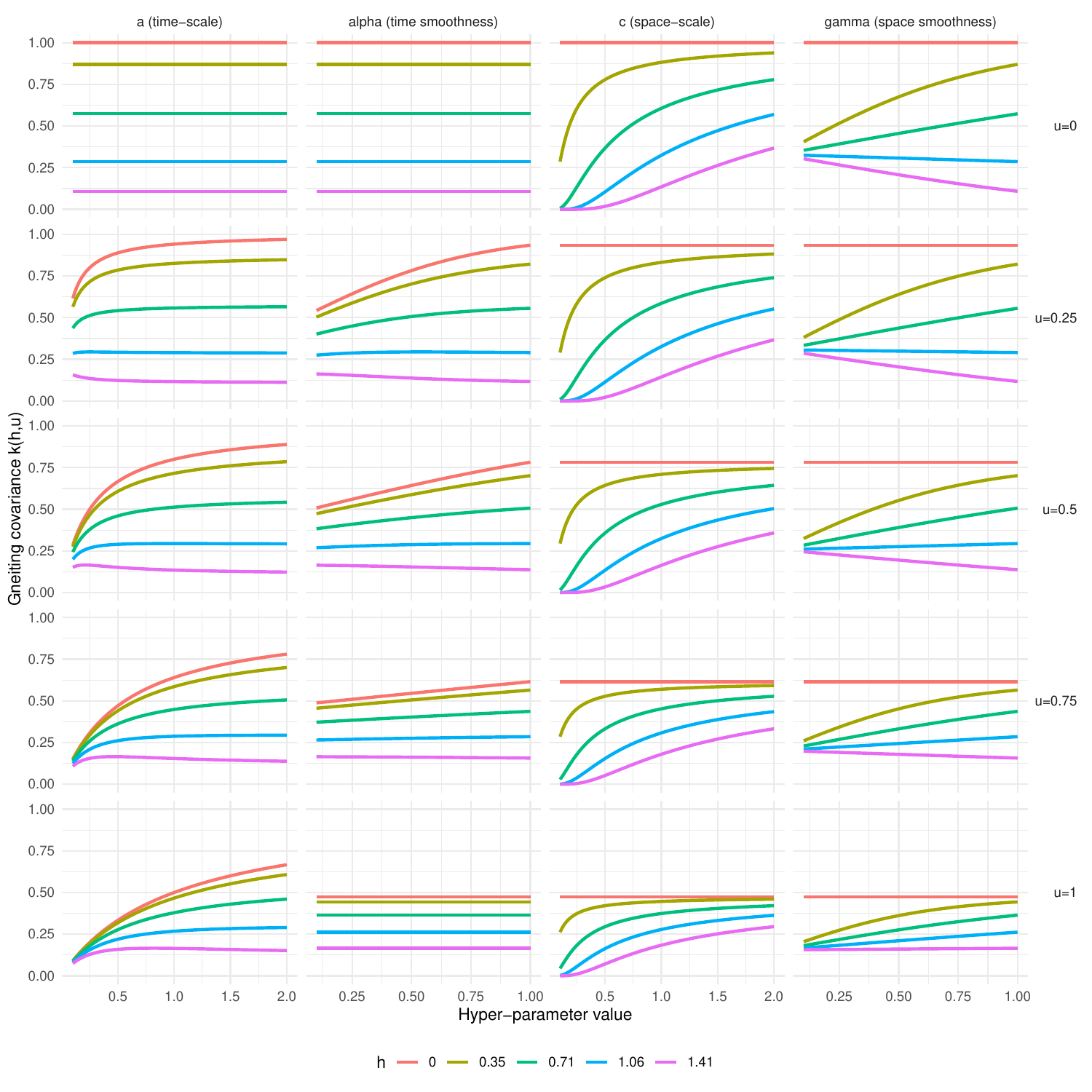}
    \caption{Sensitivity of the Gneiting covariance to its four hyper-parameters: each panel (left to right) shows how varying the temporal lengthscale, temporal smoothness, spatial lengthscale, and spatial smoothness affects the covariance curve for several fixed spatial distances $h$ (colored lines) across multiple time lags $u$.}
    \label{fig:gneiting_param}
\end{figure}

\begin{table}[H]
\centering

\caption{R-hat values for each model. R-hat values close to one suggest convergence.} \label{tab:rhat}

\begin{tabular}{lr}
  \hline
Model & R-hat \\ 
  \hline
Separable ICM-RBF & 1.00 \\ 
  Separable RBF-RBF & 1.00 \\ 
  Nonseparable Gneiting & 1.00\\
   \hline
\end{tabular}

\end{table}

\begin{figure}[H]
  \centering

  \begin{subfigure}[t]{0.29\textwidth}
    \centering
    \includegraphics[width=\linewidth]{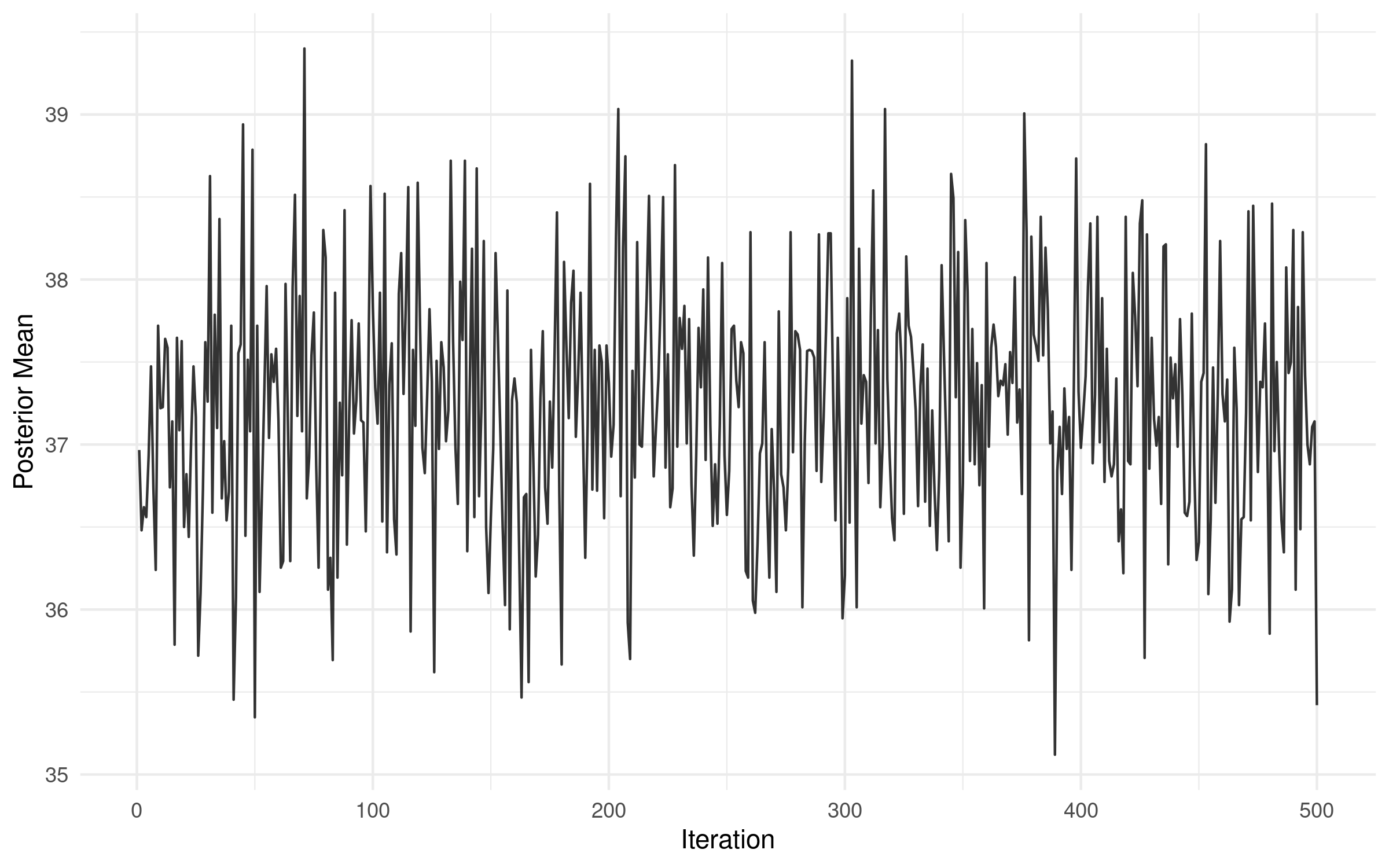}
    \caption{Separable ICM-RBF Kernel GP Model}
    \label{fig:TP_ICM}
  \end{subfigure}
  \hfill
  \begin{subfigure}[t]{0.29\textwidth}
    \centering
    \includegraphics[width=\linewidth]{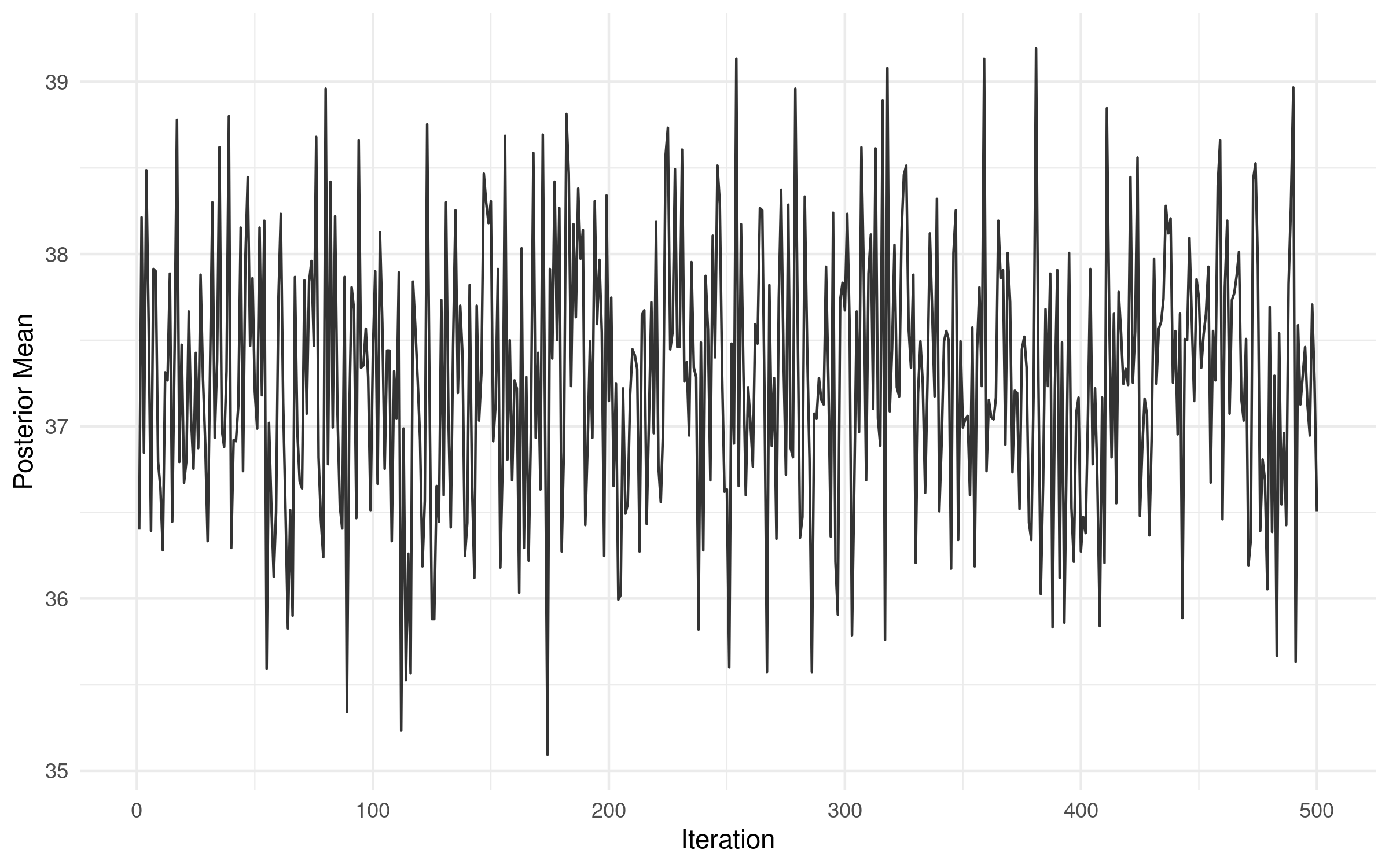}
    \caption{Separable RBF-RBF Kernel GP Model}
    \label{fig:TP_sep}
  \end{subfigure}
  \hfill
  \begin{subfigure}[t]{0.29\textwidth}
    \centering
    \includegraphics[width=\linewidth]{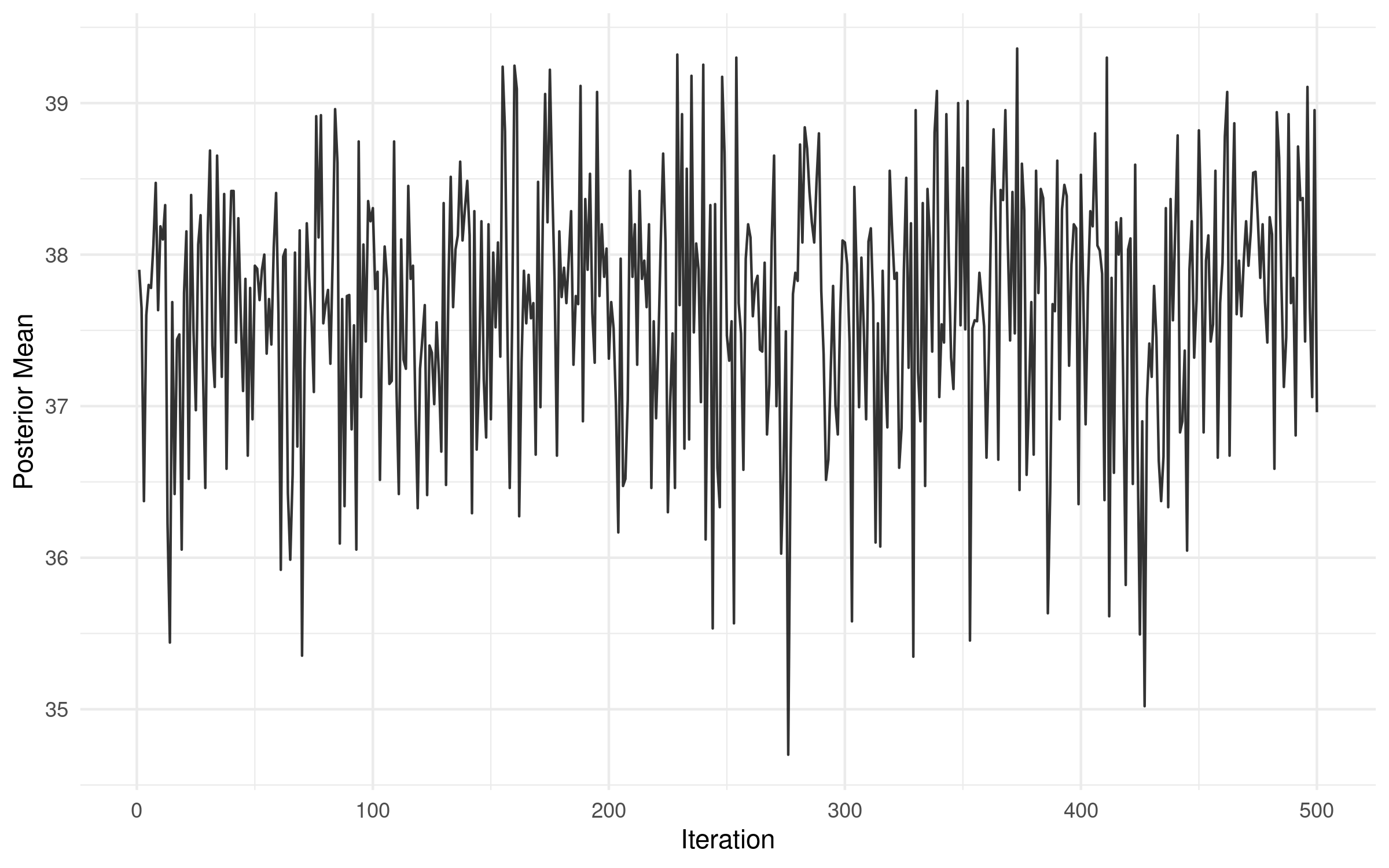}
    \caption{Nonseparable Gneiting Kernel GP Model}
    \label{fig:TP_nonsep}
  \end{subfigure}

  \caption{
Trace plots of the posterior mean across treated counties at time point $t=10$. 
Panel (a) shows results from the Separable ICM-RBF Kernel GP model, panel (b) shows results from the Separable RBF-RBF Kernel GP model, and panel (c) shows results from the Nonseparable Gneiting Kernel GP model. All plots demonstrate stable sampling behavior of the average latent function value over MCMC iterations.
}

  \label{fig:trace_plots}
\end{figure}

\begin{figure}
    \centering
    \includegraphics[width=1\linewidth]{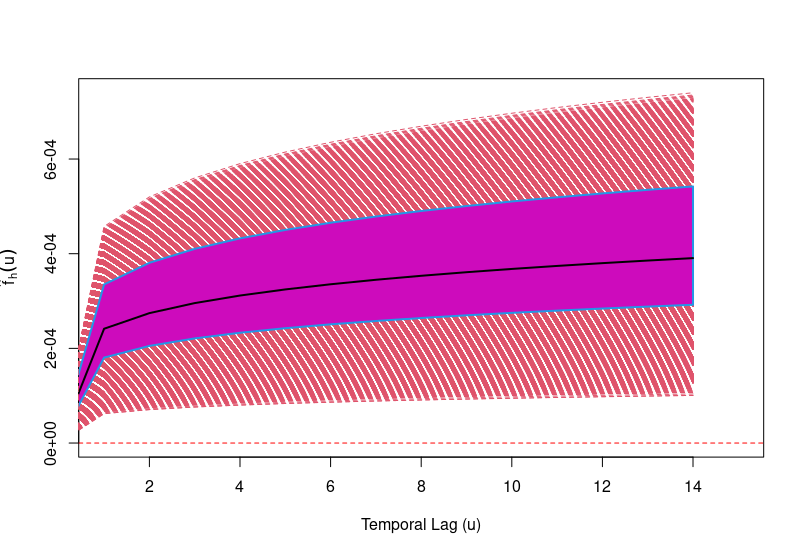}
    \caption{Estimated functional boxplots of the separability test function for the Katrina mortality dataset. In this plots, the functional median is shown as a central line, the 50 \% central region as a shaded band, and any extreme curves as outliers.}
    \label{fig:funcbox_katrina}
\end{figure}

\end{document}